\documentclass[twocolumn]{aastex62}
\usepackage{amsmath}
\usepackage{appendix}

\newcommand{\msun}{\textrm{M}_\odot}
\newcommand{\kpc}{\textrm{kpc}}

\usepackage{lineno}

\usepackage{booktabs}
\usepackage{comment}

\renewcommand{\tablename}{Table}

\usepackage{xcolor}
\newcommand{\YC}[1]{{\color{violet} \bf YC: #1}}

\shorttitle{Missing GC streams}
\shortauthors{Pearson et al.}

\defcitealias{Chen2023}{CG23}
\defcitealias{chengnedin2022}{CG22}

\begin{document}\sloppy\sloppypar\raggedbottom\frenchspacing 

\title{Forecasting the Population of Globular Cluster Streams in  Milky Way-type Galaxies}

 \author[0000-0003-0256-5446]{Sarah Pearson}
\affiliation{Niels Bohr International Academy \& DARK, Niels Bohr Institute, University of Copenhagen, Blegdamsvej 17, 2100 Copenhagen,  Denmark}
 \email{sarah.pearson@nbi.ku.dk}
 \correspondingauthor{Sarah Pearson}

\author[0000-0002-7846-9787]{Ana Bonaca}
\affiliation{The Observatories of the Carnegie Institution for Science, 813 Santa Barbara Street, Pasadena, CA 91101, USA}

\author[0000-0002-5970-2563]{Yingtian Chen}
\affiliation{Department of Astronomy, University of Michigan, Ann Arbor, MI 48109, USA}

\author[0000-0001-9852-9954]{Oleg~Y.~Gnedin}
\affiliation{Department of Astronomy, University of Michigan, Ann Arbor, MI 48109, USA}

\begin{abstract}\noindent 
Thin stellar streams originating from globular clusters are among the most sensitive tracers of low-mass dark-matter subhalos.
Joint analysis of the entire population of stellar streams will place the most robust constraints on the dark-matter subhalo mass function, and therefore the nature of dark matter.
Here we use a hierarchical 
model of globular cluster formation to forecast the total number, masses and radial distribution of dissolved globular cluster in Milky Way-like galaxies.
Furthermore, we generate mock stellar streams from these progenitors' orbital histories taking into account the clusters' formation and accretion time, mass, and metallicity.
Out of $\sim$10,000 clusters more massive than $10^4\,\msun$, $\sim$9000 dissolved in the central bulge and are fully phase-mixed at the present, while the remaining $\sim$1000 survive as coherent stellar streams.
This suggests that the current census of $\sim$80 globular cluster streams in the Milky Way is severely incomplete. 
Beyond 15 kpc from the Galactic center we are missing $\sim$100 streams, of which the vast majority are from accreted GCs. 
Deep Rubin photometry $(g\lesssim27.5)$ would be able to detect these streams, even the most distant ones beyond $> 75\,\kpc$.
We also find that M31 will have an abundance of streams at galactocentric radii of 30-100 kpc. 
We conclude that future surveys will find a multitude 
of stellar streams from globular clusters which can be used for dark matter subhalo searches.   
\end{abstract}

\keywords{Globular star clusters (656), Galaxy dynamics(591), Dark matter(353), Stellar streams(2166), Galaxy structure(622), Galaxy dark matter halos(1880), Galaxy stellar halos(598)}

\section{Introduction} \label{sec:intro}

In the Milky Way (MW), we know of $\sim$170 globular clusters \citep[GCs;][]{Harris1996,vasiliev2021} and $\sim$80 stellar streams emerging from GCs or from fully disrupted GCs \citep{Mateu2023}. It is unclear, however, if this is a complete sample of the MW's GC stellar stream population, or if we are missing a fraction of the stellar streams due to current observational limitations. 

Stellar streams from GCs form when stars leave the cluster near the tidal radius of the GC and continue their orbit around the Galaxy while leading or trailing the cluster. Lower mass GCs or more extended GCs are more susceptible to tides from the underlying potential and will disrupt more easily. Particularly clusters on orbits that bring them close to the galactic center are more susceptible to tidal stripping, while streams from GCs at larger orbital radii might not disrupt into streams, since they experience a weaker tidal field from the Galaxy. From our vantage point in the MW, however, we are limited by foreground contamination and simultaneously biased towards finding brighter streams that are closer to the Sun. It is therefore an open question which streams are missing from our current sample, and where such streams reside in the Galaxy.

Most of the observed MW GC streams have present day stellar masses ranging from $10^3$ to $3\times10^4~ {\rm M}_{\odot}$ \citep{Shipp2018}, although some have been detected with only 100s of stars \citep[e.g.,][]{malhan2022}. 
\citet{Cerny2022} discovered very faint star clusters in the MW, which could be remnants of tidally disrupted low mass clusters. However, most of the observed streams in the MW likely originated from globular clusters with initial masses of $10^4 - 5\times10^6~ {\rm M}_{\odot}$ \citep[see e.g.,][]{Shipp2018,Patrick2022}. 
The MW GC stellar streams are located within a galactocentric radius of $R_{\rm Gal} < 35$ kpc \citep{Mateu2023}, with the exception of one stream, the Eridanus stream emerging from an old GC, located at $R_{\rm Gal} \sim 95$ kpc \citep{Mateu2023,Myeong2017}. 
Since the first GC stellar stream discoveries in the Milky Way \citep[e.g.,][]{grillmair1995}, most stellar streams have been detected with photometry and matched filtering techniques \citep[e.g.,][]{odenkirchen2001,rockosi2002,bonaca2012, koposov2014,bernard2014,Shipp2018}. Thanks to the {\it Gaia} Mission \citep{Gaia2016,Gaia2018,Gaia2021}, we have now detected several more streams through a combination of  individual stars location, kinematics and photometry  (see \citealt{Mateu2023} and specific examples in \citealt{Malhan2018,ibata2018,bianchini2019,Malhan2021,jensen2021, malhan2022}).

Finding more streams, particularly in the outskirts of the Milky Way, is important in the quest for the nature of the dark matter particle. According to $\Lambda$CDM, there should be a large population of low-mass subhalos in the MW \citep[e.g.,][]{Boehm2014}, and gravitational interactions between stellar streams and subhalos can leave behind underdensities (gaps) in the streams \citep[e.g.,][]{yoon2011, erkal2017,pricewhelan2018,deboer2018,Bonaca2020,tavangar2022}. Other prescriptions of the nature of the dark matter particle predict that subhalos with masses lower than $10^6$ M$_{\odot}$ should not exist \citep[e.g.,][]{Bullock2017}.
In the inner parts of the Galaxy, streams can be perturbed by both the bar \citep{hattori2016a,pricewhelan2016b,erkal2017,Pearson2017,Bonaca2020}, molecular clouds \citep{Amorisco2016}, and spiral arms \citep{Banik2019}, all of which can lead to similar gap signatures in the streams as produced by interactions with dark matter subhalos. The Galactic outskirts is an ideal location to study gaps in streams, because streams at large galactocentric radii will have experienced less disturbances from baryonic perturbers and subhalos will have experienced less tidal disruption \citep[e.g.,][]{garrison2017,nadler2021}.  

For external galaxies, \citet{Pearson2019} predicted that the Nancy Grace Roman Space Telescope ({\it Roman}) will be able to detect thin streams from GCs in external galaxies out to $\sim$6.2 Mpc, which includes 493 galaxies. 
Additionally, \citet{aganze2023} recently showed that {\it Roman} can detect gaps in these thin streams out to $2-3$ Mpc. This opens up an entirely new era of statistical analyses of gap characteristics in thin stellar streams, which can help constrain dark matter models. 
With automated stream searches \citep[e.g.,][]{Pearson2022,shih2022,shih2024}, stellar streams at larger galactocentric radii should be easier to detect in external galaxies due to the greater contrasts with the stellar halos at large galactocentric distances. However, before we plan future observations, we need to know whether streams form at large galactocentric radii, and how many streams we should expect to find per galaxy. 

In this paper, we investigate whether there is a population of undetected GC streams at large galactocentric radii in MW type galaxies. 
We track the fully disrupted  GC from the GC catalogs based on the \citet[][hereafter CG22]{chengnedin2022} and \citet[][hereafter CG23]{Chen2023} models to investigate where we should expect to find stellar streams at present day. We analyze the fully disrupted clusters' statistical properties, run mock stream simulations of their GC progenitors from the time of disruption until present day, and estimate the observability of these streams for Rubin-type photometry and magnitude limits \citep{ivezic2008,laine2018,wagner2019}.

The paper is organized as follows: in Section \ref{sec:sims} we describe the \citetalias{chengnedin2022} and \citetalias{Chen2023} catalogs, in Section \ref{sec:methods} we describe how we generate mock streams and synthetic photometry for the stream stars, in Section \ref{sec:results} we describe the results of our analyses of the catalogs and mock streams. We discuss these results in Section \ref{sec:disc}, and conclude in Section \ref{sec:concl}.

\section{Model Sample of Globular Clusters}
\label{sec:sims}

GCs are collection of gravitationally bound stars typically formed more than $10$~Gyr ago. These bright objects are extremely compact and formed in less than a few Myr. Direct simulations of cluster formation \citep[e.g.,][]{howard2018,li2019,ma2020,chen2021,grudic2021} require sub-parsec spatial resolution and can only run for $\sim$10~Myr. To model the entire GC population in a MW-mass galaxy, one has to make simplifications by either running simulations with sub-grid prescriptions for cluster formation \citep[e.g.,][]{li2017,reina2022} or post-process existing cosmological simulations to analytically determine the properties of model GCs \citep[e.g.,][]{kruijssen2008,maratov2010,li2014,renaud2017,Choksi2018,creasey2019,Phipps2020,Halbesma2020,Valenzuela2021}. See also  \citet{rodriguez2023} for an approach including collisional $N$-body modeling of star clusters.

In this paper, we use the post-processing model presented in \citetalias{chengnedin2022} and \citetalias{Chen2023}, which  incorporates key processes controlling the formation and evolution of individual GCs. Such a model can efficiently output mock catalogs of model GCs\footnote{\url{https://github.com/ognedin/gc_model_mw}} with parameters selected to reproduce the distributions of key GC properties, including mass, age, metallicity, positions, and velocities (\citetalias{chengnedin2022}; \citetalias{Chen2023}; \citealt{chen2024}). These outputs serve as the inputs for the subsequent stream modeling throughout this paper. 

The catalogs incorporate an analytical model of GC formation and evolution on carefully selected galaxies in the hydrodynamic simulation Illustris TNG50-1 \citep[][hereafter TNG50]{Nelson2019}.
The model has a detailed prescription of tidal disruption considering the location-dependent tidal field. The tidal disruption shapes the initial cluster mass function to the present version in close agreement with observations (see Fig.~1 of \citetalias{Chen2023}). Therefore, this model successfully links the birth masses of GCs to the present mass and provides reliable dissolution time for the disrupted GCs, which is essential for self-consistently modeling streams as GC debris.

The model includes 4 steps: 1) formation of GCs; 2) cluster sampling; 3) particle assignment; and 4) tidal disruption of GCs. 
The first step triggers GC formation when the parent galaxy encounters a major merger or strong accretion. This is quantified by its mass growth rate exceeding a threshold value. We then use observed galactic scaling relations to compute the gas mass and metallicity of the parent galaxy. We assume the newly-formed clusters' total mass is proportional to the cold gas mass \citep{Kravtsov2005}, and their metallicity is directly inherited from the parent galaxy. 

In the cluster sampling step, we sample the mass of individual clusters using a \citet{Schechter76} initial cluster mass function, with a high-mass end cutoff at $10^7\ \msun$. In the \citetalias{chengnedin2022} model the initial cluster mass was populated down to $10^5\ \msun$. In \citetalias{Chen2023} and this work, we follow the initial cluster mass function down to $10^4\ \msun$. A lower minimal mass is needed as we are interested in whether any of the low-mass GC streams on orbits with large pericenters could survive until present day. We do not model clusters below initial mass $10^4\ \msun$ as they tend to dissolve quickly ($<1$~Gyr) due to tidal disruption. In these first two steps, the only simulation input is the halo merger tree. We define the GCs formed in the main host branch as \textit{in-situ} clusters, and the other GCs originally formed in satellite galaxies as \textit{ex-situ} clusters. 

Next, we assign collisionless simulation particles as tracers of the newly-formed GCs to obtain positional and kinematic information. For hydrodynamic simulations like TNG50, we preferentially select young stellar particles formed around the time of GC formation in the model. For $\lesssim10\%$ of model clusters we cannot find enough stellar particle candidates and use dark matter particles instead. We track the trajectories of the selected tracer particles to obtain the present-day positions and velocities of model GCs. 

Finally, we compute the mass loss due to the stellar evolution and tidal disruption. The latter is modeled with an updated prescription as in \citetalias{Chen2023}, where the disruption time scale is proportional to the cluster mass and inversely proportional to the tidal frequency. The tidal frequency is an effective measure of the strength of the tidal field along cluster orbits, and is calculated using the eigenvalues of tidal tensor, which is numerically evaluated as the second-order finite differences of the gravitational potential.

We apply the model to several carefully selected galaxies whose properties are similar to the MW. We require the MW analogs to have 1) total mass within the range $10^{11.9-12.3}\ \msun$, 2) maximum circular velocity between $210$ and $270\ {\rm km\,s^{-1}}$ \citep{eilers2019}, 3) at least one major merger (mass ratio $>1/4$) between $10-12$~Gyr ago, representing the Gaia-Sausage/Enceladus \citep{helmi2018,belokurov2018}, 4) no major merger in the last 10~Gyr, and 5) formed $25-35\%$ of their present-day stellar mass at 10~Gyr ago \citep{leitner2012}. There are seven TNG50 galaxies satisfying all criteria above. {We focus on the model which best matches the observed GC system in terms of total number, mass function, spatial and metallicity distributions, and velocity dispersion. We refer to this model as MW1. \citet{pillepich2023} included MW1 in their MW-M31 catalog, and \citet{semenov2024} included MW1 as one of the five earliest spin-up galaxies with a well-settled extended disk at $z=0$.
Note that there are two other matches which meet the criteria above. We refer to these as MW2 and MW3. However, these TNG50 MW analogs do not have extended disks at $z=0$, and we therefore only use these as a measure of systematic uncertainty for our results. The details of each potential are outlined in Appendix \ref{sec:analytic_potentials}.

The model has three adjustable parameters: the threshold to trigger a GC formation event, the proportionality factor between the galaxy cold gas mass and the mass of newly-formed clusters, and the correction for the underestimate of the numerically-calculated tidal tensor.  The profiles and concentrations of the GCs are not included explicitly, since the tidal radius scale of GCs (20-50pc) is beyond the spatial resolution of TNG50, but the tidal stripping formalism does incorporate information about the overall tidal field, and adopts the mass loss rate from the parametrization in \citet{Gieles2023}. \citet{Gieles2023} results were based on several direct N-body simulations of star clusters with different density profiles and concentrations, including possible effects of retained stellar-mass black holes, which can accelerate mass loss near the final stages of disruption. 
We calibrate the three adjustable parameters by comparing the statistics of model GCs in the MW analogs to the real MW GCs. The statistics include total cluster number, mass function, metallicity distribution, radial distribution, and velocity dispersion. 
The model assumes that all GCs in the MW with $m > 10^4~ {\rm M}_{\odot}$ have been discovered. If a vast population of new GCs are detected with future missions, (e.g., the Nancy Grace Roman Space Telescope, Euclid, or the Vera C. Rubin Observatory) our models need to be re-calibrated. As our analysis throughout this paper only focuses on streams from fully disrupted GCs, our results will not be affected by the discovery of a few more GCs.

We note that the catalog used here is different from the most recent model catalogs for MW and M31 published in \citet{chen2024}. The latter incorporated a slightly different galaxy mass--metallicity relation and a different dependence of tidal disruption timescale on the initial and current GC mass. Also, \citet{chen2024} modeled the scatter of galactic scaling relations using a Gaussian process. However, these small modifications do not notably change final GC properties. Moreover, for the purpose of this paper, none of the previously mentioned prescriptions is directly related to the modeling of streams. In addition to the different model prescriptions, \citet{chen2024} fit model parameters jointly for the MW and M31 systems, while here we only optimize the parameters for the MW galaxy. We emphasize that both catalogs can well match the observations of MW GCs.

In this paper we are interested in tracking fully disrupted globular clusters, i.e. stellar streams. Since the model GC catalog populations are well matched to the MW population of observed GCs, we assume that the catalog streams should also be well-matched to the MW's population of observed and unobserved stellar streams. However, the insufficient spatial and temporal resolution in TNG50 snapshots may lead to imperfect predictions. We discuss the limitations of using the catalogs for stream predictions in Section~\ref{sec:discgcform}.

\section{Stream modeling methods}\label{sec:methods}

To test which fully disrupted clusters from our catalogs can be identified as stellar streams today, we generate mock stellar stream models.
We place the stream progenitors on orbits in a Milky Way-like potential (\S\ref{sec:potential}), generate stream models using the particle-spray method (\S\ref{sec:mocks}), and assign stellar masses and synthetic photometry to stream particles to assess their observability (\S\ref{sec:photometry}).

\subsection{Gravitational potential}
\label{sec:potential}

To calculate tidal disruption of their GC populations, \citetalias{chengnedin2022} represented the host galaxy potential with a basis function expansion (BFE).
Their model accurately represents the present-day gravitational potential of the TNG50 halos, however, it is computationally expensive to evaluate -- prohibitively so for generating mock stellar streams with the complete census of star particles (see \S\ref{sec:mocks}).
We therefore adopt a more idealized, but faster to evaluate, model of a Milky Way-like gravitational potential.

We generate mock streams in a three-component model, consisting of a \citet{hernquist:1990} bulge, a \citet{mn:1975} disk, and a \citet{nfw:1996} halo, fit to reproduce the host galaxy potential.
The fitting procedure, as well as the exact analytical forms and the best-fit parameters for each of the host halos are described in Appendix~\ref{sec:analytic_potentials}.
In Appendix Figure~\ref{fig:BFE} we compare the BFE and analytic circular velocity profiles for the three hosts.
Overall, the analytic potential model recovers the circular velocity at a 10\%-level.
While the cluster orbits differ in detail, the orbital peri- and apocenters are similar in the two potential models.
The median offset and the standard deviation in both the pericenters and apocenters for the entire cluster population are smaller than 1\,\kpc.
This indicates the stream forecasts generated in our adopted potential to be reasonably realistic.
Note that we do not incorporate the effects of time-evolving potentials in this work, and solely fit the potentials to the $z=0$ TNG50 snapshots. While time-dependence of the halo and  time-dependence of the rotating Galactic bar at the center of the Galaxy can perturb the morphologies of stellar streams and affect stream observabilities \citep[e.g.,][]{hattori2016a,pricewhelan2016b, erkal2019,shipp2021,Lilleengen2022,dillamore2022,thomas2023,arora2023}, we leave such investigations for future work.

\subsection{Generating mock streams}
\label{sec:mocks}

To estimate how many stellar streams in the Milky Way can be observed, we created a particle-spray model \citep{Fardal2015} of each globular cluster stream in our sample using \texttt{MockStreamGenerator} in the \texttt{gala} package \citep{gala2017,gala2020}.
Starting from the position at which a cluster would be at the present had it not dissolved, we integrated their orbits back in time to the start of their dissolution process in the Milky Way. 

\begin{figure}
\centering
\includegraphics[width=\columnwidth]{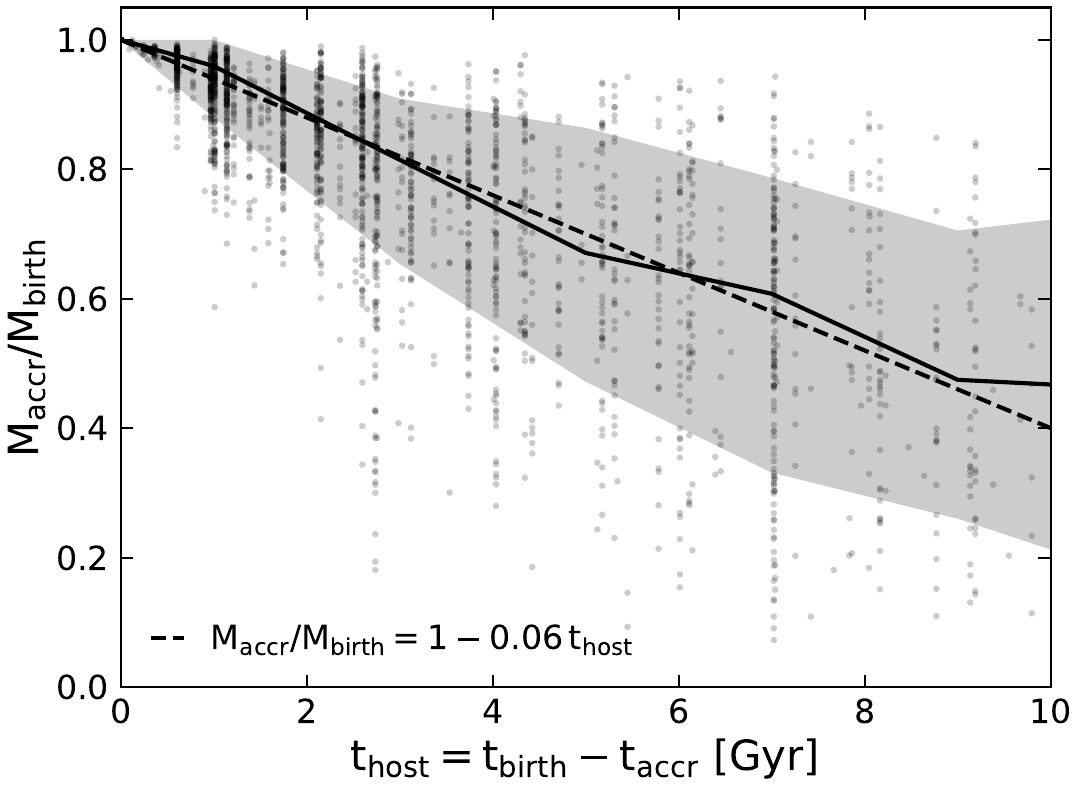}
\caption{Ratio of cluster mass at the time of accretion and at birth as a function of time spent in the host dwarf galaxy prior to accretion onto the central Milky Way-like halo, for all accreted clusters in the model catalog. Cluster mass decreases approximately linearly with the time spent in the original host galaxy. The solid line shows the median trends, while the shaded region shows the $16-84$th percentiles.}
\label{fig:m_accrete}
\end{figure}

Because we simulate only streams expected to form in the Milky Way, the in-situ and ex-situ clusters are initialized differently.
For in-situ clusters the initial position is their position at birth, but ex-situ clusters are initialized at the time of accretion onto the Milky Way.
Furthermore, we account for mass that ex-situ clusters lost before accretion.
Figure~\ref{fig:m_accrete} shows the ratio of the cluster mass at the accretion time and at birth as a function of time spent in the host dwarf galaxy prior to accretion.
There is a significant scatter for all model clusters, however, the mass at accretion, $M_{\rm accr}$, is on average well described by relation:
\begin{equation}\label{eq:accrete}
  \frac{M_{\rm accr}}{M_{\rm birth}} \approx 1 - 0.06 \left(\frac{t_{\rm host}}{\rm{Gyr}}\right)
\end{equation}
where $M_{\rm birth}$ is the initial cluster mass, and $t_{\rm host}$ the age of the cluster when it accreted onto the Milky Way-like host.
Stars unbound from globular clusters in the original dwarf galaxy progenitor are expected to form diffuse tidal debris in the Milky Way, often with complex morphology \citep{carlberg:2020, malhan:2021}.
Capturing these features would require generating mock streams in a time-evolving potential, which we leave to future work.
As a result, our models of ex-situ streams constitute only their stars stripped after accretion, and we use Equation~\ref{eq:accrete} to estimate the cluster masses at the time of accretion. 

After initializing the cluster, we integrate its orbit forward and at every time step release tracer particles, setting their (cluster mass-dependent) escape positions and velocities following \citet{Fardal2015}. 
We release stars at the two Lagrange points uniformly in time every 0.5\,Myr. In the next section we discuss how we determine and assign the total number of stars to each stream. 
We stop generating new particles at the cluster dissolution time, and continue evolving the completely dissolved model stream until the present.
The cluster formation, accretion, and dissolution times, as well as their birth masses are taken directly from the catalog of the GC population, making the generated set of streams a self-consistent extension to the model.

\subsection{Synthetic photometry for stream stars}
\label{sec:photometry}

To determine what fraction of our stream population would be detectable with current and upcoming observational facilities, we need to apply observational limitations, the dominant of which is a flux limit.
In this section we describe how we produced synthetic photometry for each star particle in our stream models based on the catalog values of each cluster's metallicity, age, and initial mass (or mass at accretion for ex-situ clusters).
We first estimated stellar masses of stream stars, and then determined their present-day luminosities using stellar isochrones.

We assumed a \citet{kroupa:2001} stellar initial mass function, and for a given cluster mass, sampled it between $0.1\,\msun$ and $15\,\msun$.
This is a standard assumption in dynamical modeling of GCs \citep[e.g.,][]{baumgardt:2003}. 
Next, we used MIST isochrones \citep{dotter:2016, choi:2016} of the appropriate age and metallicity for a given cluster taken directly from the catalog to determine the highest mass star still luminous today.
We then generated a stream model with the number of star particles matching the expected number of less massive, luminous stars surviving until present day for each  cluster of a specific mass, age, and metallicity. 
Our $10^4\,\msun$ stream models have on average $\approx12,000$ members, $10^5\,\msun$ have $\approx60,000$, and $10^6\,\msun$ have $\approx700,000$ stars.

In our stream modeling, we assume that stars get tidally stripped in the order of mass, with the low mass stars being released first and high mass stars last.
While idealized, this procedure captures a general trend observed in dynamical models of GC evolution: more massive stars sink deeper into the potential well due to mass segregation and take longer to evaporate \citep[e.g.,][]{henon:1969, vesperini:1997, baumgardt:2003}.
As a result, the oldest parts of stellar streams are predominantly populated with low-mass stars and harder to detect \citep{balbinot:2018}.

To determine luminosity of a given stream star particle, we used isochrones from the MIST project \citep{dotter:2016, choi:2016}.
Progenitor cluster metallicities and ages provided in our GC catalog are largely continuous, but for practical reasons we used a grid spaced by 0.1\,dex in metallicity and 0.1\,Gyr in age.
For an isochrone matching the cluster's age and metallicity, we interpolated the synthetic absolute magnitudes as a function of stellar mass using a univariate spline.
We report synthetic photometry in the LSST $g,r$ bands to make forecasts for the Vera C. Rubin Observatory.
To convert absolute magnitudes to apparent magnitudes as observable by LSST, we adopt the \texttt{astropy} V.4.0 Galactocentric reference frame, a right-handed Cartesian reference frame in which the Sun is at a distance of 8.122\,kpc from the Galactic center, and 20.8\,pc above the Galactic plane.

\section{Results}\label{sec:results}

In this Section, we first show the results of our analysis of the statistical properties of the fully disrupted GCs from the Milky Way catalog (MW1, Section \ref{sec:summarystat}). We subsequently investigate which streams survive until present day (Section \ref{sec:survivingstreams}) and determine their observability (Section \ref{sec:resobsstreams}).

\subsection{GC streams in the model catalog}
\label{sec:summarystat}

 \begin{figure}
    \centering
    \includegraphics[width=\columnwidth]{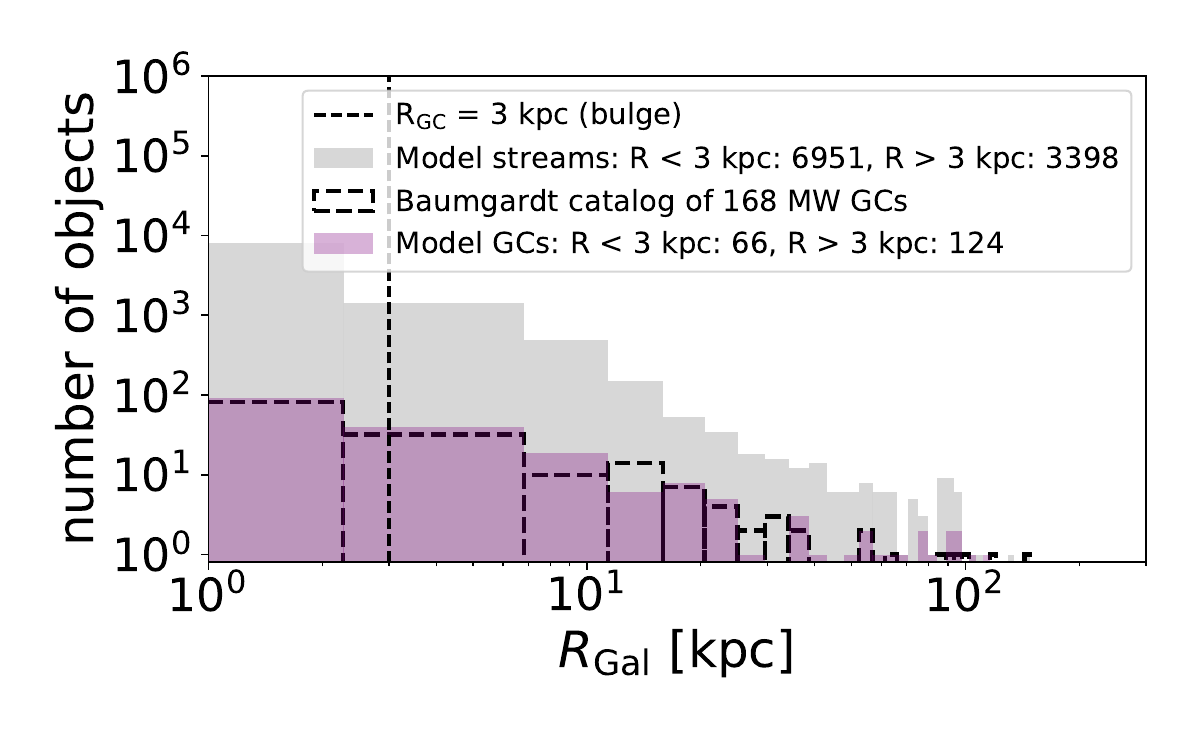}
    \caption{The number of GCs (purple) and fully disrupted GCs (gray) from the  MW1 catalog as a function of galactocentric radius, binned in widths of 4.5~kpc. There are 190 GCs total while there are 10349 fully disrupted GCs. The dashed vertical lines shows $R_{\rm Gal} = 3$~kpc, and the labels note how many objects are located within and beyond this radius, which is roughly the size of the bulge. The dashed histogram shows the 168 observed MW GCs from the updated \citet{baumgardt2019} catalog.}
    \label{fig:numberofobjects}
\end{figure}

In Figure \ref{fig:numberofobjects} we show the population of GCs (purple) and fully disrupted GCs (stream candidates, gray) from the MW1 catalog as a function of galactocentric radius.  
There are a total of 190 surviving GCs in the catalog, 
which is higher than the 170 observed GCs reported by \citet{vasiliev2021}, but similar to recent findings by \citet{garro2024}, who reported 200 GCs. The histogram outlined as a black dashed line shows a recent compilation of 168 observed MW GCs updated from \citet{baumgardt2019}. The MW1 catalog and the observed MW GCs follow a similar radial distribution (\citetalias{chengnedin2022}, \citetalias{Chen2023}). 
The farthest model GC is at a galactocentric radius of ${\rm R}_{GC} = 118$ kpc (note that its orbital apocenter location could be larger than this present day galactocentric location). The vertical dashed line shows the approximate radius of the bulge (${\rm R}_{GC} = 3$ kpc), and we find that 66 of the model GCs reside within this region, while 124 are located beyond the bulge.

The fully disrupted GCs, which are stellar stream candidates (gray), extend beyond ${\rm R}_{\rm Gal} > 100$  kpc, and there are more than 50 times as many fully disrupted GCs than surviving GCs in the catalog. 6951 fully disrupted GCs are located within the bulge region, many of which are likely impossible to detect at present day\footnote{Note that with the use of kinematic information in the search criteria, in recent years \texttt{STREAMFINDER} \citep{Malhan2018,Malhan2021} has  been finding some streams in the inner MW.}, while 3398 fully disrupted objects reside beyond the bulge region. 

\begin{figure}
    \centering
    \includegraphics[width=\columnwidth]{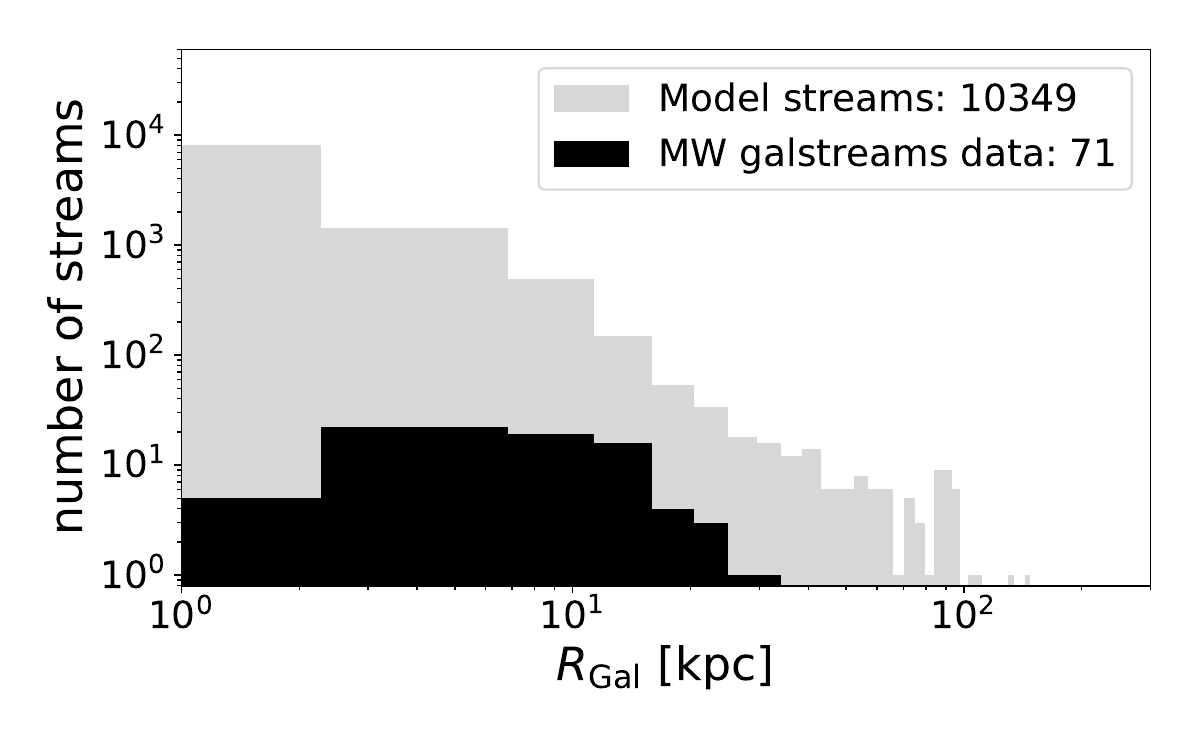}
    \caption{The gray histogram shows the galactocentric radius of each fully disrupted GC from the MW catalog (same as in Figure~\ref{fig:numberofobjects}). Note that many of these stream candidates will not be observable at present day. The black histogram shows the mean galactocentric distance for all observed fully disrupted GC streams in the MW calculated based on the stream track information provided in \texttt{Galstreams} (\citealt{Mateu2023}), where we have excluded dwarf streams (9 total) and stellar streams emerging from known GCs (10 total).}
    \label{fig:galstreams}
\end{figure}

We compare the GC stream population predictions from the catalog to the observed population of streams in the Milky Way in Figure \ref{fig:galstreams}. To carry out this comparison, we use the \citet{Mateu2023} \texttt{Galstreams} package, which contains information on the tracks of all observed streams in the Milky Way. 
To provide a fair comparison to the model GC stream population, we first exclude all streams in the \texttt{Galstreams} which are associated with surviving globular clusters (12 total\footnote{Pal 5, Pal 15, Eridanus, NGC 3201-Gjöll, M92, M5, M68-Fjörm, NGC 288, NGC 2298, NGC 5466, M2 (NGC 7089), and $\omega$Cen-Fimbulthul.}), and we exclude the dwarf galaxy streams in the sample (9 total\footnote{Sgr, LSM-1, Orphan-Chenab, Cetus, Elqui, Indus, Jhelum, Palca, Tucana III.}). After these cuts, there are 71 observed fully disrupted GC streams in the MW sample.

\begin{figure}
    \centering
    \includegraphics[width=\columnwidth]{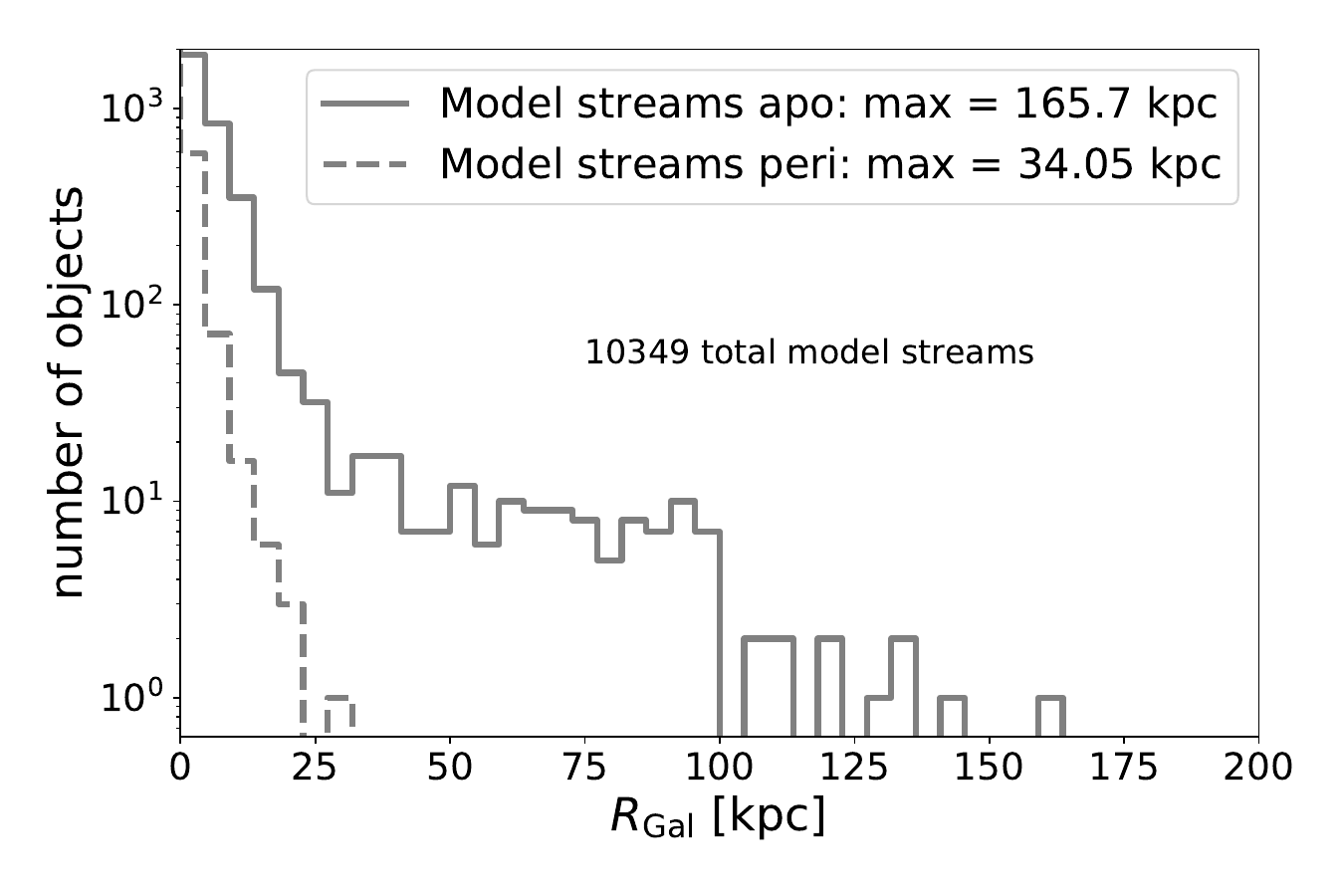}
    \caption{The apocenter distribution (solid gray) and pericenter distribution (dashed gray) of all fully disrupted clusters (stream candidates) from the MW1 GC catalog.}
    \label{fig:apoperi}
\end{figure}

We use \texttt{astropy} \citep{astropy13,astropy18} to transform all stream tracks into galactocentric coordinates. For each stream we then compute the median galactocentric radius. We plot the  distribution of the observed stream locations in Figure \ref{fig:galstreams} (black bars).  The stream located at $R_{\rm Gal}\approx 80$ kpc is the Eridanus stream \citep{Myeong2017}. Most observed streams in the MW reside within 30 kpc of the galactic center. 

\begin{figure}
    \centering
    \includegraphics[width=\columnwidth]{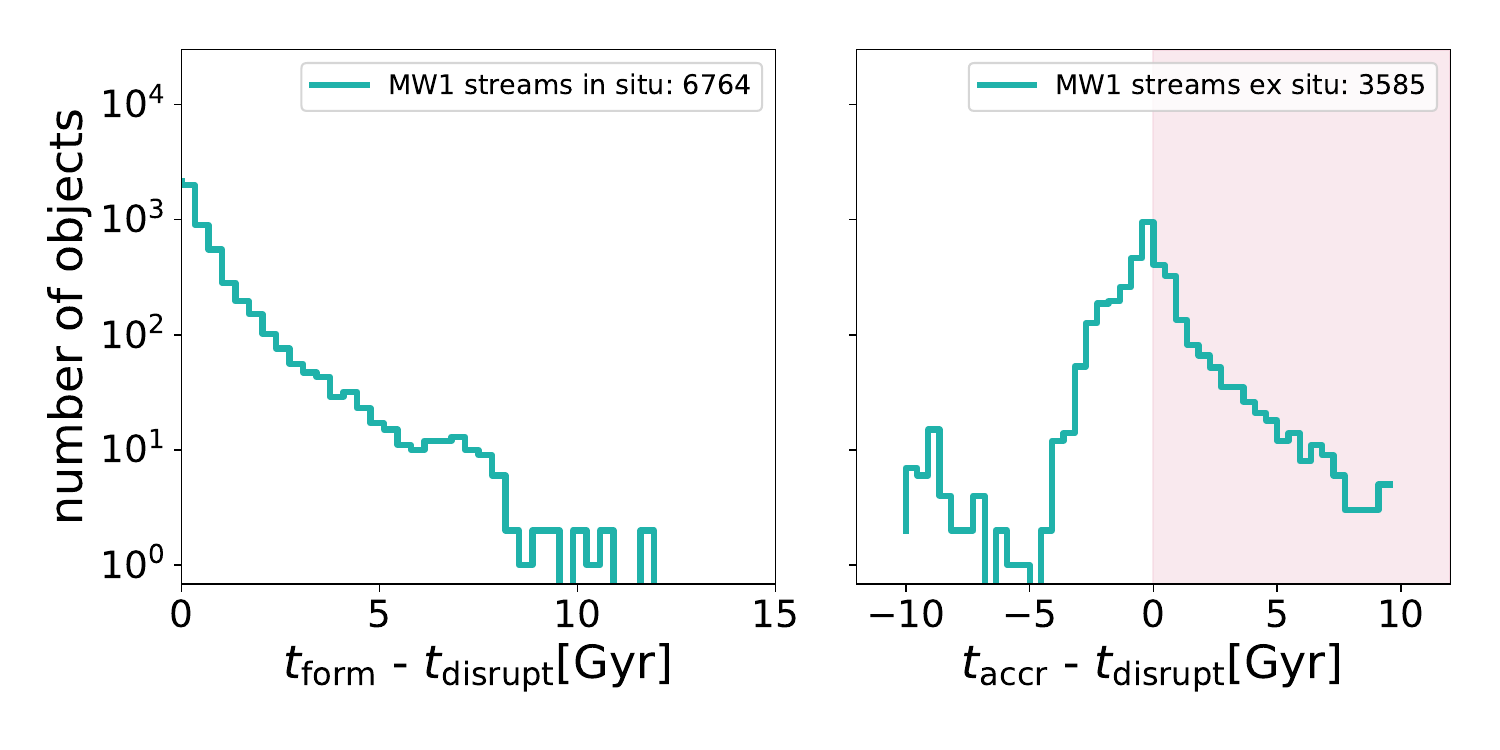}
    \caption{{\it Left:} The number of fully disrupted GCs which were formed in situ from the MW1 catalog vs the time it takes for each of the objects to fully disrupt after their formation. {\it Right:} The number of fully disrupted ex situ (accreted) GCs in the MW1 catalog vs the time it takes to fully disrupt each object after accretion. If this value is negative, the cluster was fully disrupted prior to accretion. The shaded region shows the objects, which we include in our further analysis of the stream candidates.}
    \label{fig:accretion}
\end{figure}

In Figure~\ref{fig:galstreams} we see that there is a large population of stream candidates in the catalog (gray) compared to the discovered GC streams in the MW (black), and a large population of catalog stream candidates are located at  $R_{\rm Gal}> 30$ kpc. In Section \ref{sec:resobsstreams}, we explore whether some of the catalog stream candidates can survive until present day and whether they can be observed.

In Figure~\ref{fig:apoperi} we show the distribution of apocenters (gray solid line) and pericenters (gray dashed line) for the 10349 model stream candidates. Streams spend most time on their orbits close to apocenter, and for future prospects of discovering thin GC streams in the outskirts of galaxies, it is encouraging that some fully disrupted model GCs  have large apocenters. There could be a large population of stellar streams at a galactocentric radii beyond 50 kpc. 

To determine whether the fully disrupted GCs will survive as streams until present day, we need to further investigate the model stream progenitors' birth masses, the time since the stream progenitors were accreted or formed, and the time since the stream progenitors were fully disrupted. It is  more likely to observe a stream at present day if the GC was disrupting after accretion, if the pericenter is large, and if the GC fully disrupted more recently \citep[see e.g.,][]{balbinot:2018,Gieles2021,roberts2024}. 

In the left panel of Figure~\ref{fig:accretion} we plot the number of objects which were formed in-situ and eventually fully disrupted in the MW1 catalog (i.e. the stream candidates) as a function of how long it takes each of the objects to disrupt after the formation of the cluster. 
While most in-situ streams disrupt quite rapidly after the GC formation, some objects fully disrupt close to the present day.

In the right panel of Figure~\ref{fig:accretion} we show all of the objects in the MW1 model catalog, which were accreted and eventually fully disrupted (i.e. the stream candidates) as a function of how long it took these objects to fully disrupt after accretion. If this value is negative, the objects fully disrupted prior to accretion. We do not include these objects in our analysis of stream survival and observability in the next sections. For the objects that disrupt after accretion (shaded region), we use the GC mass at the time of accretion (see Equation \ref{eq:accrete}), when we model the streams' evolution. Some of the objects fully disrupt close to present day.

\subsection{Surviving model streams} \label{sec:survivingstreams}

\begin{figure*}
	\includegraphics[width=\textwidth]{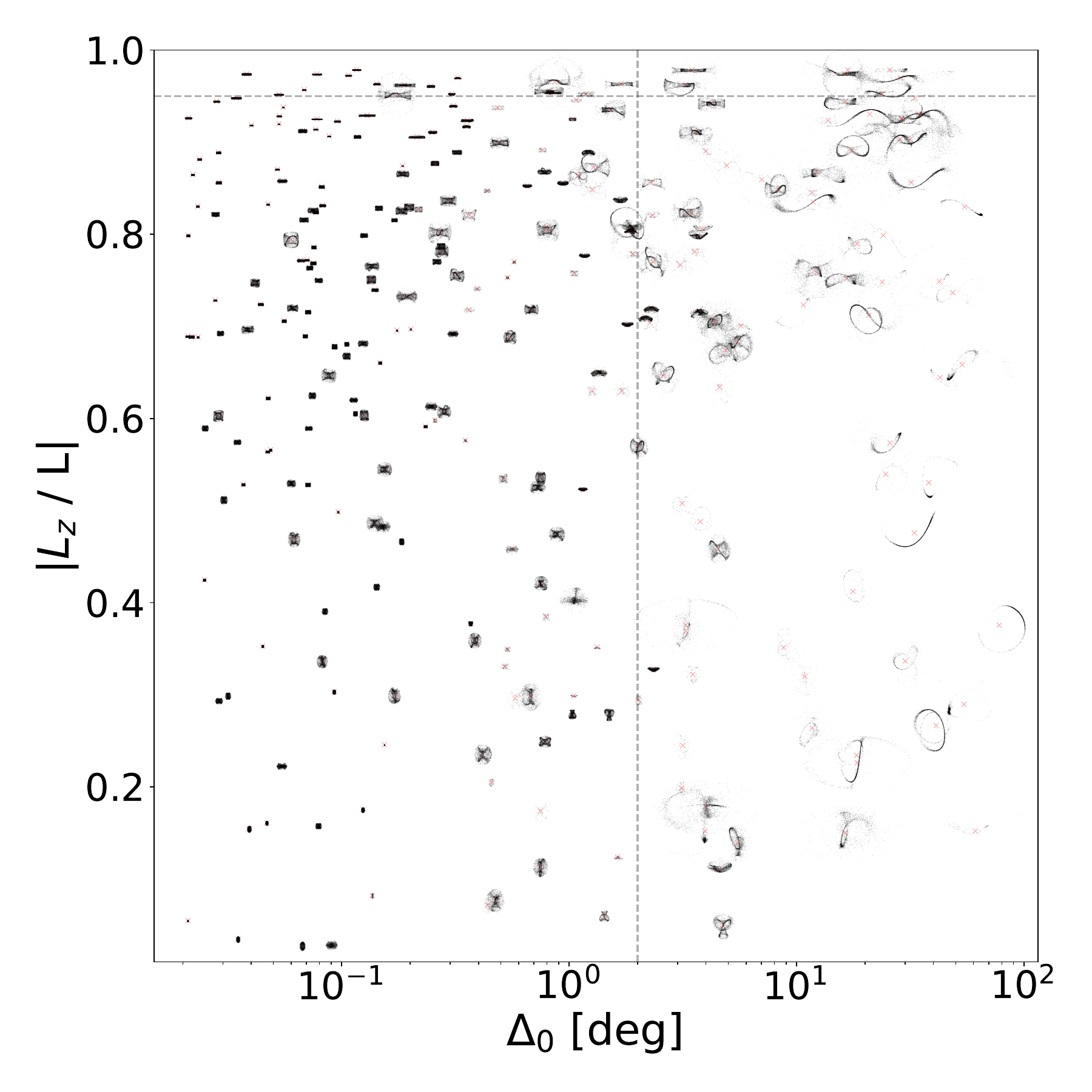}
\caption{The absolute value of the $z$-fraction of the angular momentum vs the streaminess parameter for a subset of 300 of the fully disrupted objects in the MW1 catalog. For each object in $|L_z/L|$ vs $\Delta_0$-space, we over plot the model stream in galactic coordinates as a visualization of our criterion. The dashed lines indicate the boundaries $|L_z/L| = 0.95$ and $\Delta_0 = 2$~deg. The bottom right quadrant shows fully disrupted objects which survive as streams until present day, based on our selection criterion.}
    \label{fig:streaminess}
\end{figure*}

To investigate which of the fully disrupted model GCs survive as coherent stellar streams until present day, and how those streams are distributed, we simulate the evolution of the fully disrupted clusters as described in Section~\ref{sec:mocks}. Note that the amount of stars in all model streams have been drawn from a GC initial mass function based on their progenitor cluster's birth mass, and that each stream's evolution is modeled based on the GC orbit, GC time of formation or accretion (for ex situ clusters), and time of dissolution. To mimic the effects of mass segregation in clusters, the lowest mass stars are lost from the cluster first. Once the progenitor is fully disrupted, we do not add more stars to the stream, but let the stream evolve and phase mix in time.

We model each of the disrupted objects that did not fully dissolve prior to accretion (8994 total objects for MW1)
in their respective three-component potential (see Appendix \ref{sec:analytic_potentials}). For MW1, 10 objects failed in the mock stream generation, as their orbits came too close to the Galactic center. 
The pericenters of each of the failed orbits are all $<0.1$ kpc, and they have the $z$-component of the angular momentum close to 0. We show a summary of these failed orbits in Appendix~\ref{sec:failedstreams}.

For each mock stream, 
we transform the stream member positions into galactic latitude and longitude coordinates and compute the following ``streaminess'' parameter:
\begin{equation}\label{eq:streaminess}
    \Delta_0 = \sqrt{{\rm med}(b)^2  + {\rm med}(l)^2}
\end{equation}
where {\rm med}($b$) is the median of all galactocentric latitudes for that particular stream's members, and  {\rm med}($l$) is the median of all galactocentric longitudes for that particular stream's members. 
If $\Delta_0 \approx 0$, the debris is either dispersed close the center of the galaxy (i.e. in the bulge region) or fully phase mixed symmetrically around the center. 
For each stream, we also calculate the absolute value of the fraction of the total angular momentum contained in the $z$-component, $|L_z/L|$. 
Streams with $|L_z/L| \approx 1$ solely rotate within the disk, which makes them hard to detect in the Milky Way.

In Figure~\ref{fig:streaminess}, we illustrate the distributions of the $z$-fraction of the angular momentum $|L_z/L|$ vs the streaminess parameter $\Delta_0$ for a subset of 300 of our model streams from the 8984 modeled objects from the MW1 catalog.
For each object, we visualize its stars in galactic coordinates at the location corresponding to the streams' $|L_z/L|$ vs $\Delta_0$ value. 
Through visual inspection, we determine that streams with $\Delta_0<2$~deg or $|L_z/L| > 0.95$ either appear fully phase mixed or are in the disk of the galaxy. We classify these as ``debris''. 
The lower right quadrant of the black dashed lines represents the streams which we classify as ``streams surviving until present day''. Thus our streaminess criterion is: 
\begin{equation}
    \Delta_0 > 2\,\mathrm{deg} \quad\mathrm{and}\quad |L_z/L| < 0.95
\end{equation}
From this analysis, there are 1814 surviving streams at present day from MW1.

\begin{figure}
	\includegraphics[width=\columnwidth] {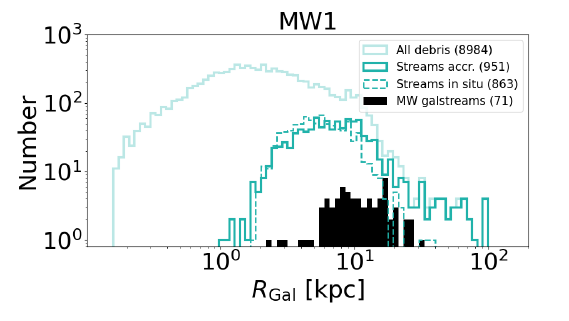}
\caption{Galactocentric radius of all disrupted debris in the model catalog (light solid line), all accreted streams surviving until present day (dark solid line), and all streams surviving until present day from in situ formed GCs (dashed line) for the MW1 catalog. Streams are defined as the objects which survived the streaminess criterion from  Figure \ref{fig:streaminess}. The black histogram shows the known fully disrupted GCs in the MW from \texttt{Galstreams}. There are more than 20 times as many predicted as detected streams, and most of the streams at large $R_{\rm Gal}$ are from accreted objects.}
    \label{fig:rgal_all}
\end{figure}

In Figure~\ref{fig:rgal_all}, we show the median galactocentric radius calculated for each star in each of the fully disrupted objects classified as ``debris'' (solid light line). For all objects that were classified as surviving streams at present day, we plot the in situ disrupted GCs as a dashed line, and the accreted disrupted GCs as a solid dark line. Most of the surviving streams are located beyond the bulge region due to the streaminess criterion. We again compare this to the observed fully disrupted GC streams in the MW (black histogram). We summarize the amount of surviving model streams predicted to be located at various galactocentric radii to the observed MW streams in the same radius bins in Table~\ref{tab:streams}. We list both galactocentric and heliocentric distances of all predictied model streams and of the observed MW streams.
Note that we do not include a Galactic bar when we model streams in this work, which could affect the observability of some of the stellar streams in the inner regions \citep[e.g.,][]{pricewhelan2016b,hattori2016a,thomas2023}. Additionally, we do not incorporate effects from a live dark matter halo, such as accretion events, which could perturb stream in the outer galaxy and affect their observability \citep{erkal2019,shipp2021,Lilleengen2022,dillamore2022,arora2023}.

The streams which survive until present day at galactocentric radii $R_{\rm Gal} >15$~kpc are predominantly from accreted GCs.  
These objects might be most useful in our quest for interactions with dark matter subhalos, as they are far from the baryonic perturbers (e.g. spiral arms, bar, molecular clouds) in the inner Galaxy. However, accreted GC streams may also exhibit complex morphologies due to pre-processing in their respective dwarf host halos \citep{malhan:2021}.

\begin{deluxetable}{ccccc}
\tablecaption{Summary of how many surviving GC streams we expect in the Milky Way at various galactocentric radii.}
\tablecolumns{8}
\tablenum{1}\label{tab:streams}
\tablewidth{0pt}
\tablehead{\colhead{\bf $D$} & 
\colhead{\bf MW1}  &{\bf MW obs } & \colhead{\bf MW1} &{\bf MW obs }  \\
\colhead{[kpc]} & \colhead{from center} &\colhead{from center} & \colhead{from Sun} &  \colhead{from Sun}  \\
\colhead{} & 
\colhead{\# streams } & \colhead{\# streams} & \colhead{\# streams} & \colhead{\# streams}
}
\startdata
$0-15$ & 1691  &  51 &1561& 42 \\
$16-30$  & 81 &  19& 203&26  \\
$31-45$  & 20  &  1& 27& 2 \\
$45-60$  & 9  &  0 &9 & 1\\
$61-75$  & 7   &  0&9& 0 \\
$>75$  & 6 &  0&5 &0 \\
\hline
{\bf all}  &  1814 & 71 & 1814& 71  \\
\enddata
\tablenotetext{}{ Column (1) distance bins calculated based on the median galactocentric radius of each stream member for a given stream. Column (2) number of surviving model streams in each radius bin calculated as from the center of the galaxy, Column (3) number of observed fully disrupted MW GC streams from Galstreams in each radius bin calculated as from the center of the galaxy. Column (4) same as Columns 2, but now the model streams have been converted to ICRS coordinates, and the distance is from the Sun. Column (5) Same as column 4 but for the MW Galstreams observations of fully disrupted GCs.}
\end{deluxetable}

The total stellar mass of the 8984 fully disrupted clusters, which did not fully disrupt prior to accretion from the MW1 model catalog, is $\sim 4.3\times10^8$~M$_{\odot}$, and the total stellar mass of the surviving GC model streams is $\sim 6.4\times10^7$~M$_{\odot}$. This is comparable to the estimated mass lost by the MW's surviving globular cluster population in the field in \citet{ferrone2023}. \citet{Deason2019} used red giant branch (RGB) stars selected from Gaia data release 2 out 100 kpc, to estimate the mass of the MW stellar halo and found M$_{\rm *,halo}\sim 1.4\times10^9$~M$_{\odot}$. This mass estimate includes debris from dwarf accretion events, and is a factor of $\sim 3$ larger than the mass of the disrupted model GCs. For comparison, the total stellar mass of the bulge of the MW is M$_{\rm *,bulge}\sim 2\times10^{10}$ M$_{\odot}$ \citep{valenti2016}. As expected, the stellar mass brought in from dwarf accretion events is crucial to explain both the stellar halo mass and bulge mass of the MW (e.g., debris from the GES merger, \citealt{helmi2018,belokurov2018}).

In Figure~\ref{fig:rgal_gcs}, we plot the ratio of all streams which survive until present day to the surviving GCs as a function of galactocentric radius. For this analysis, we include the three MW catalogs (MW1, MW2, and MW3) introduced in Section \ref{sec:sims}, and discussed in Appendix \ref{sec:analytic_potentials}. From these two catalogs, we again generated mock streams for the fully disrupted objects, which were not fully disrupted prior to accretion, and determined their survivability with the streaminess criterion. 
While MW2 and MW3 do not host stable disks at $z=0$, their different accretion histories and surviving stream distributions can inform the systematic uncertainty for our results in the outskirts of each galaxy. Despite variations in the accretion histories between the three MW catalogs, the surviving stream distributions at present day in the outskirts of the galaxies ($R_{\rm Gal} >$ 15 kpc) are similar between all catalogs. 
Note that due to the lack of extended disks, 2115 and 1305 mock stream runs failed in the inner part of the galaxy for MW2 and MW3, respectively, as compared to only 10 failed streams in MW1  (see Appendix~\ref{sec:failedstreams}). 
For MW1 (teal line), there are about 10 times as many surviving streams to GCs in the inner galaxy. 
There are a similar number of surviving streams to surviving GCs in the outer galaxy for all three catalogs.
This information can be used as an estimate for how many streams we will find in extragalactic systems, where we know the GC distribution but have yet to detect the GC streams. If we detect more GCs with future missions, the predicted number of missing streams can be scaled from this figure. 

\begin{figure}
	\includegraphics[width=\columnwidth]{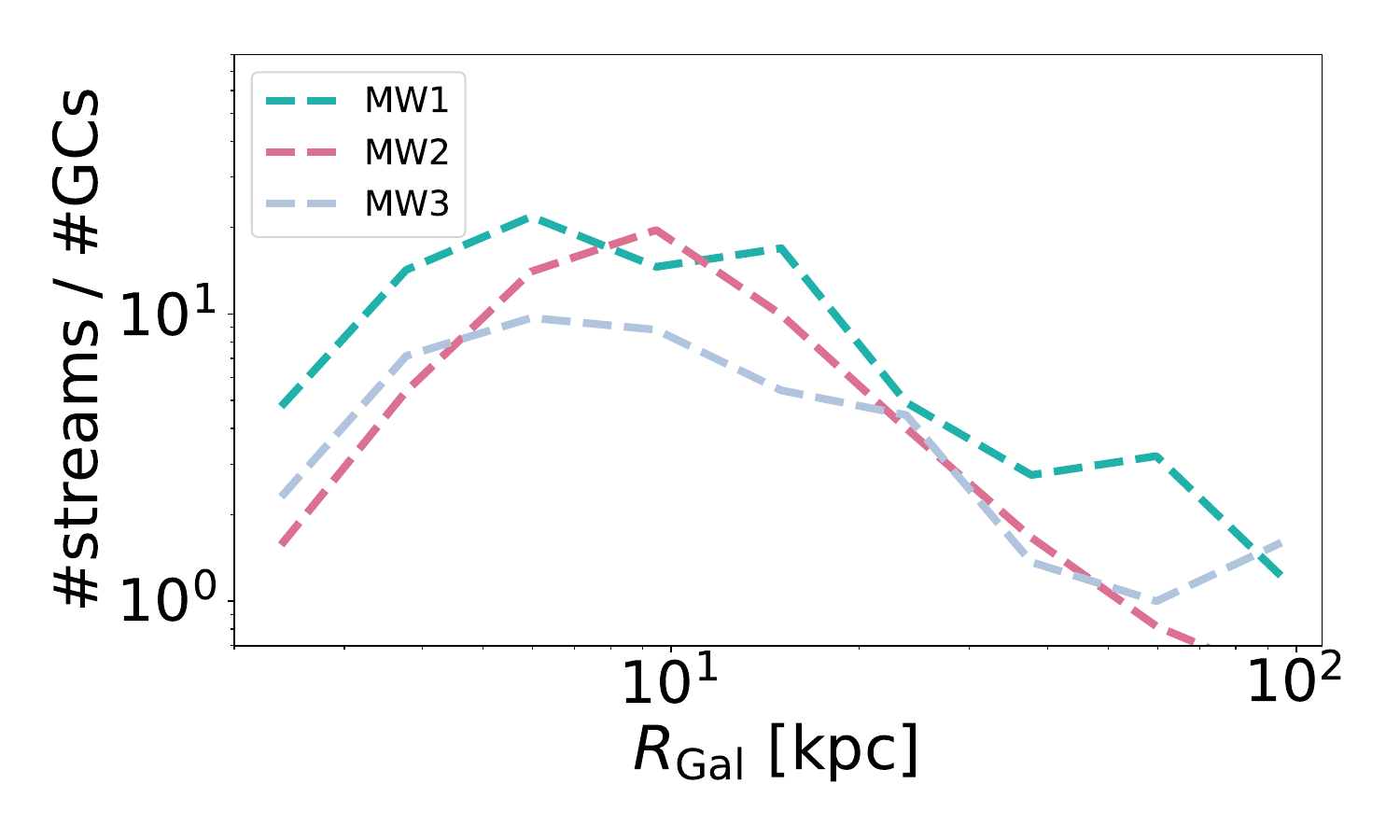}
\caption{The ratio of surviving model streams to surviving model GCs as a function of galactocentric radius for the three MW model catalogs (see Appendix~\ref{sec:analytic_potentials}). Note that the differences close to the galactic center is due to the failed mock streams in the central regions of MW2 and MW3 (see Appendix~\ref{sec:failedstreams}).
}
    \label{fig:rgal_gcs}
\end{figure}

\subsection{Observability of surviving  model streams} \label{sec:resobsstreams}
\begin{figure}
	\includegraphics[width=\columnwidth]{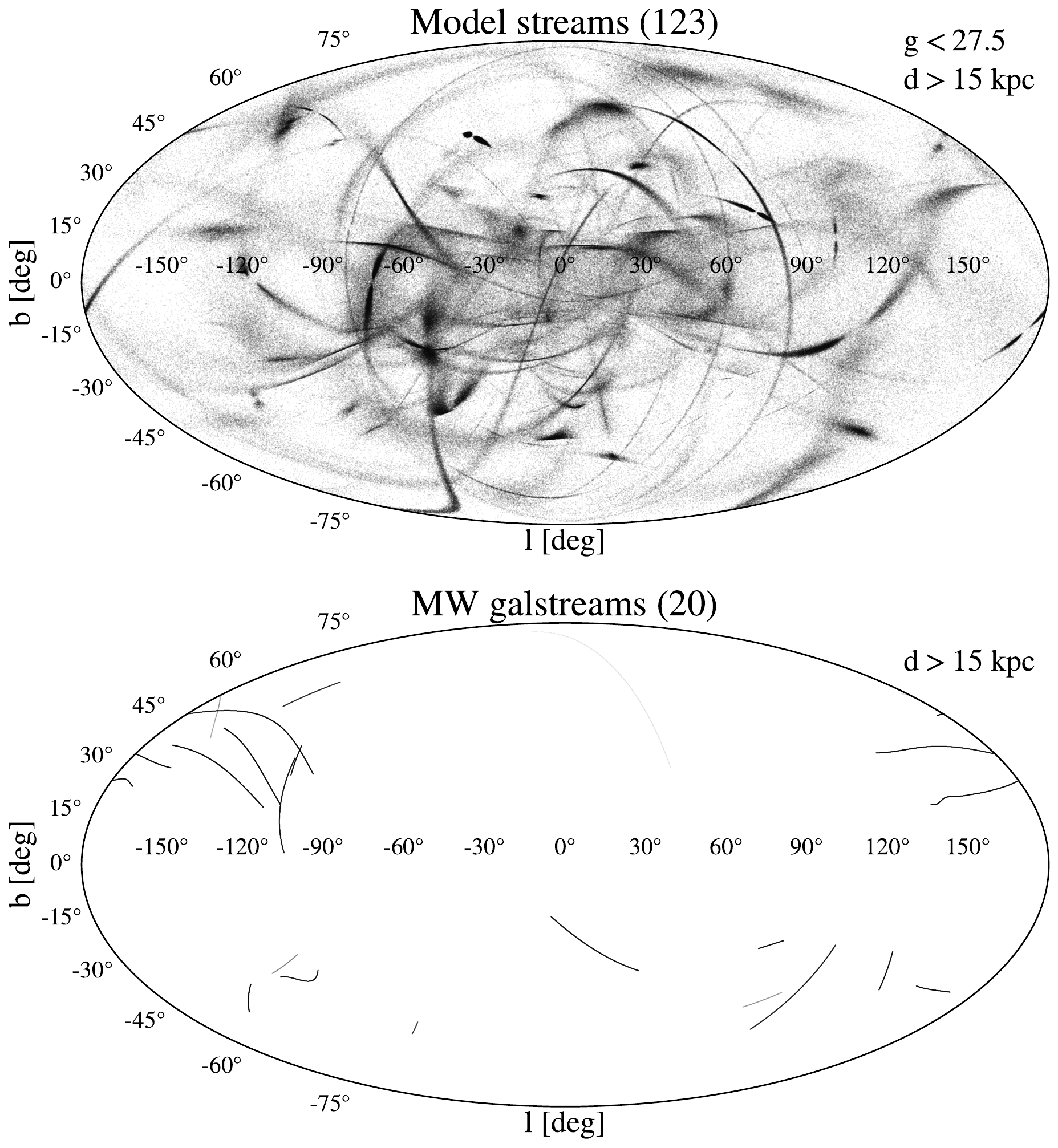}
\caption{{\bf Top:} The surviving model streams which are located at $R_{\rm Gal} >$ 15 kpc, for all stars with LSST $g$ band photometry $g<27.5$. Each of these streams have their main sequence turnoff more than one magnitude above the LSST limit for resolved stars. {\bf Bottom:} The tracks of the observed MW stream population located at $R_{\rm Gal} >$ 15 kpc from \texttt{Galstreams}. There are potentially a large population of GC streams in the MW yet to be detected at large galactocentric radii.}
    \label{fig:prplot}
\end{figure}

To determine which streams can be detected in future surveys, we investigate the observability of the model streams from MW1 with future LSST photometry. 

For a given cluster mass, age, and metallicity, we determine the stream stars' luminosities as described in Section~\ref{sec:photometry} for LSST $g,r$-band photometry.  
In the top panel of Figure~\ref{fig:prplot}, we plot all stream stars for all model streams which survived until present day, which reside beyond $R_{\rm Gal} > 15$~kpc, and which have LSST $g < 27.5$, similar to LSST limits for resolved stars \citep{LSST2009}.
This leaves 123 of the 1814 model streams from the MW1 catalog (see top panel). For each of these 123 streams, the main sequence turnoff in LSST $g,r$-bands is more than one magnitude above the LSST limits for resolved stars (all main sequence turnoffs had $g<25$). Note that we do not mock observe the streams, include a Milky Way foreground model, nor take into account the length and widths of the streams in this analysis, but since other MW streams have been detected with survey limits only one magnitude below the main sequence turnoff  \citep[e.g. the Chenab stream in][]{Shipp2018}, we argue that LSST will be able to detect these missing streams. 

To estimate stream observability in a more realistic scenario, including the impact of dust and reddening, we injected two of our distant stream models in a mock LSST catalog, specifically Data Challenge 2 produced by the Dark Energy Science Collaboration (DESC/DC2\footnote{https://github.com/LSSTDESC/DC2-production}).
DC2 is a high-latitude mock catalog that includes Milky Way foreground stars, unresolved background galaxies, a dust model, and realistic photometric uncertainties.
We found that an isochrone-based selection in the dereddened color-magnitude space easily recovers the more nearby and more massive of the streams ($R_{\rm Gal}\approx30$\,kpc, $M_{\rm GC}=10^5$\,M$_{\odot}$).
However, such a simple selection only recovers the densest parts of the more distant and less massive of the streams ($R_{\rm Gal}\approx90$\,kpc, $M_{\rm GC}=3\times10^4$\,M$_{\odot}$).
This analysis suggests that accounting for dust using dust maps will not affect stream observability outside of the Galactic plane, but that more sophisticated selections may be needed to increase the contrast of faint, distant streams (e.g., remove foreground stars using proper motions, remove background galaxies using color-color selections or potentially shapes from space-based imaging).}

\begin{figure*}
	\includegraphics[width=\textwidth]{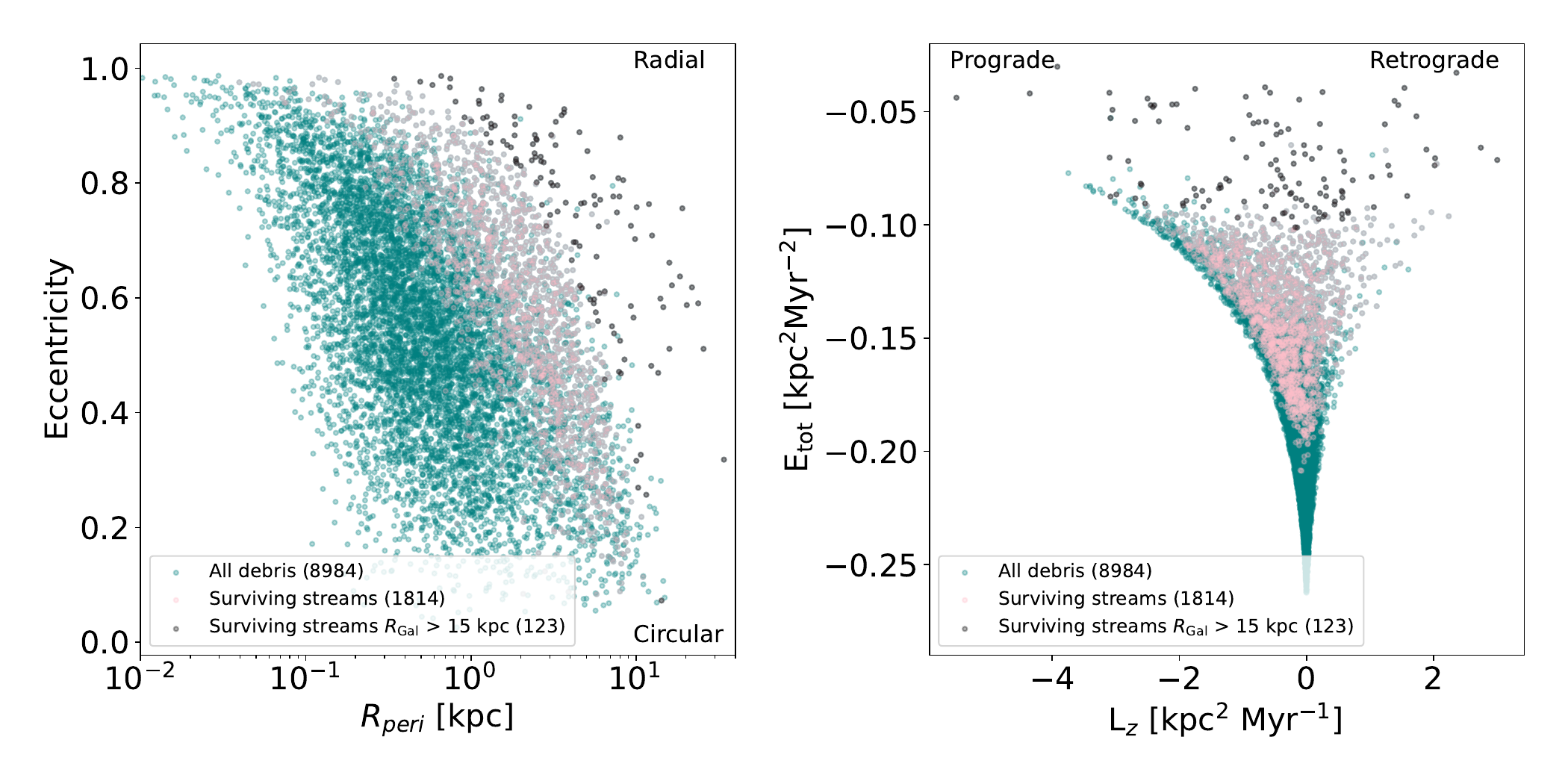}
\caption{Left: The eccentricities vs pericenter distances for all model debris (teal), for all debris which survives as streams (pink), and for all surviving model streams which have $R_{\rm Gal} >$ 15 kpc (black). The surviving model streams have higher eccentricities and pericenters than the fully phase-mixed debris. Right: total energy vs $z$ angular momentum. The surviving model streams have higher energies but random directions of motion around the galaxy. Note that we only included objects which were not fully disrupted prior to accretion (see Figure \ref{fig:accretion}).
}
    \label{fig:ecc}
\end{figure*}

For comparison, in the bottom panel of Figure~\ref{fig:prplot} we plot all known \texttt{Galstreams} GCs stream tracks from fully disrupted GCs in the MW located beyond 15~kpc \citep{Mateu2023}. We do not impose a limiting magnitude to the MW observations, as these are all detected with varying methods. Figure \ref{fig:prplot} demonstrates that LSST can detect a much larger number of streams than the currently observed population of MW GC streams. 

The average pericenters of these 123 objects is 6.1~kpc, and the average initial cluster mass of these objects is $2.9\times 10^4$~M$_{\odot}$. In the left panel of Figure \ref{fig:ecc}, we show the distribution of eccentricities vs pericentric distances for all debris (teal), for all surviving streams (pink), and for all surviving model streams beyond 15 kpc (black) from Figure \ref{fig:prplot}. The disrupted debris (teal), reaches very low pericenters as compared to the surviving streams. Surviving streams with small pericenters tend to be on more eccentric orbits. These streams spend most of their time at larger galactocentric radii, and therefore are less likely to be fully disrupted. The streams which survive at a present day at $R_{\rm Gal} >$ 15 kpc (black), are all on high eccentricity orbits, as they otherwise would not be able to tidally strip at  $R_{\rm Gal} >$ 15 kpc. Some of these streams get close enough to the Galactic center and disk, that they might be perturbed by the bar, spiral arms, or molecular clouds, while the streams with pericenters greater than 10 kpc will be ideal for subhalo searches.

In the right panel of Figure \ref{fig:ecc}, we show the total energy vs $z$-angular momentum for all debris (teal), surviving streams (pink), and surviving streams located at $R_{\rm Gal} >$ 15 kpc (black). The streams at $R_{\rm Gal} >$ 15 kpc have higher total energies given their, on average, larger orbital radii. While most objects are on prograde orbits (74\% for all debris), there is no systematic preference for retrograde or prograde orbits for the surviving streams as compared to the fully disrupted debris (66\% prograde objects for the surviving streams, and 75\% prograde objects for the surviving streams at $R_{\rm Gal} >$ 15 kpc).

To further investigate the observability of our forecasted population of MW streams, we also calculate the surface brightness (SB) of all surviving catalog streams for MW1. To calculate the SBs of the model streams, we use the \texttt{Healpix} Python package, and first define a pixel size of $da = 0.0131$ deg$^2$ (\texttt{Healpix} level 9) and calculate the flux within each pixel from all luminous stars. We also tested \texttt{Healpix} level 7 and 8, but chose level 9, as this captures both the dense part of the low mass streams and the more diffuse parts of higher mass streams. 

\begin{figure}
    	\includegraphics[width=\columnwidth]{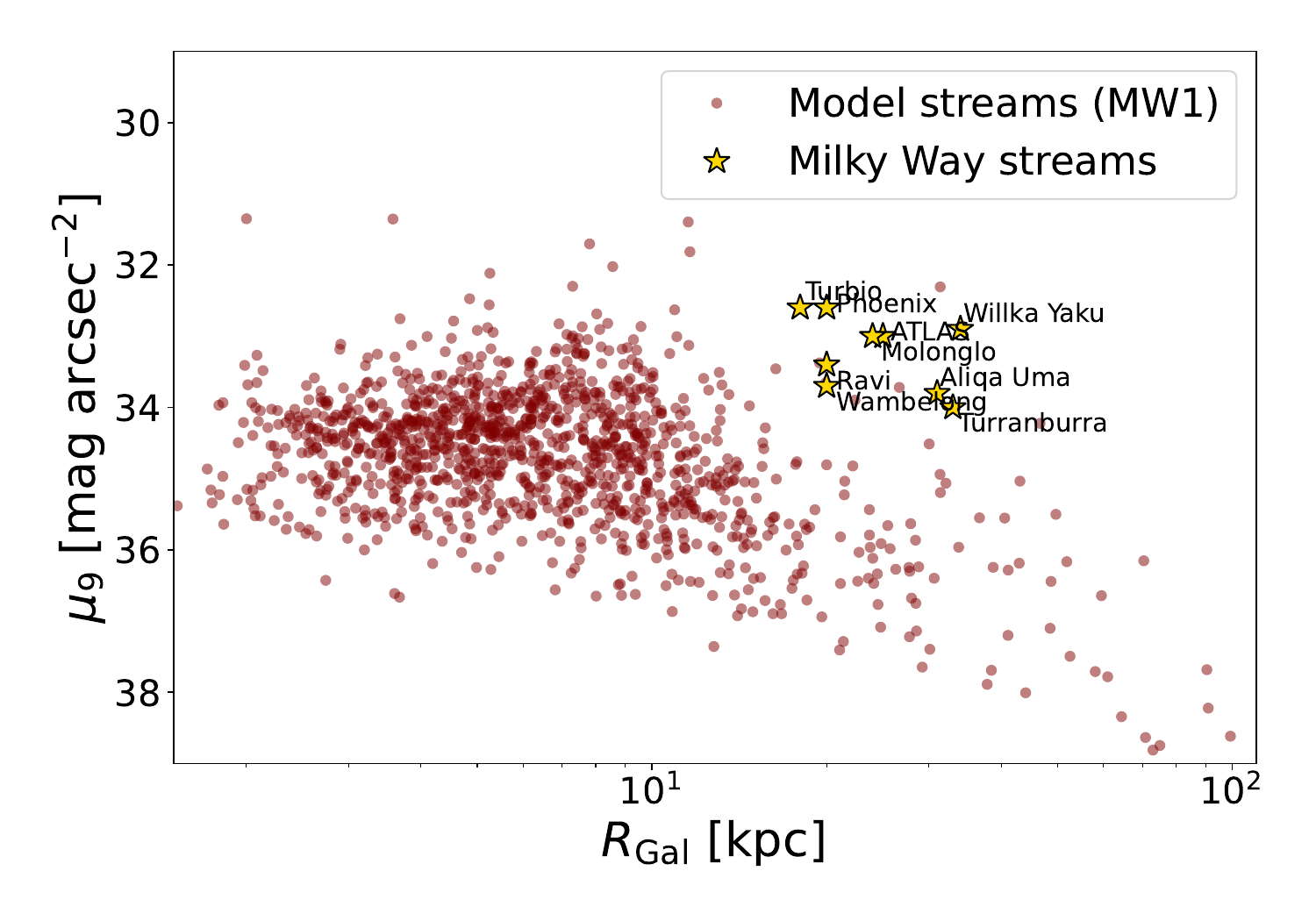}
\caption{The surface brightness for the 1814 surviving model streams from MW1 (red) and examples of observed MW GC streams (yellow stars) vs galactocentric radius. Most model streams are fainter than the MW streams, but some are similar.}
    \label{fig:rgal_SB}
\end{figure}

In Figure~\ref{fig:rgal_SB}, we show the SB of each surviving stream from the MW1 catalog as a function of galactocentric radius (red), and compare this to a subset of observed SB from \citealt{Shipp2018} (yellow stars). Note that while we calculate the flux and SB in each pixel, \citet{Shipp2018} calculated the average SB of each stream by assuming 68\% of the luminosity is contained within the observed stream width. They also assume a \citet{Chabrier2001} IMF to fit their isochrones, thus their method is different from ours. Most of the model streams are fainter (the median value is 34.7 mag/arcsec$^2$) than the observed MW streams (the median value is 33.0 mag/arcsec$^2$), several are similar. At large galactocentric radii there are a number of lower surface-brightness model streams that have evaded detection so far, and these can hopefully be discovered in deeper data. Interestingly, there should be many more lower surface-brightness streams in the inner galaxy as well.
Note also that MW streams have been observed at smaller  galactocentric radii (see MW streams in Figure~\ref{fig:rgal_all}), however we do not know the SB for these streams.

\section{Discussion} \label{sec:disc}

Our results indicate that we have only discovered about 5\% of the MW's GC stream population thus far, that hundreds of streams remain undetected within 30 kpc of the Galactic center in the Milky Way, and that tens of streams remain undetected beyond 30 kpc. Interestingly, \citet{shih2024} have reported the detection of $\approx$90 new stellar stream candidates using the \texttt{Via Machinae 2.0} algorithm and searching Gaia DR2 data \citep{Gaia2018}. Our analysis indicates that these candidates could indeed be stellar streams from CGs. 
In this Section, we discuss the limitations of our assumed stream formation and modeling efforts (Section~\ref{sec:discgcform}), future science prospects with deeper photometry of streams in the MW (Section~\ref{sec:futurescience}), and the prospects of GC stream searches in external galaxies, particularly M31 and dwarf galaxies (Section~\ref{sec:discexternal}).

\subsection{Limitations of the model GC catalogs} \label{sec:discgcform}

Throughout this work, we have simulated fully dissolved GCs to track their stellar streams and make predictions about their survivability and observability. Caveats with this method include understanding the detailed formation of globular clusters at various mass scales, their densities as birth, and cluster dissolution. 

The GC catalogs, constructed by \citetalias{chengnedin2022} and \citetalias{Chen2023} using TNG50 galaxies, are calibrated by many observational constraints at $z=0$. However, the model inherits TNG50's limited numerical resolution in space and time. The $\sim200$~Myr intervals between snapshots inadequately capture rapid accretion events that trigger cluster formation and tidal shocks. 
Our GC model instead approximately accounts for tidal disruption through an orbit-averaged correction factor based on direct N-body models \citep[see][Eq.~4]{Gieles2023}.
Spatial resolution ($\sim 0.3$~kpc) affects the dynamics of simulation particles, which may not accurately represent the actual trajectories of GCs. Moreover, some (less than 10\%) of GCs are represented by dark matter particles in TNG50 galaxies. These dynamically hotter particles have a more extended apocenter distribution due to their higher eccentricities than the stellar particles.

Additionally, as the cluster catalog was created including TNG50 dynamics, the distribution of accreted GCs captures effects of dynamical friction only on their host dwarf galaxies. 
Our mock stream simulations do not include dynamical friction, however, since the orbits are calculated in static analytic potentials. These potentials are represented by three-component models based on the present day TNG50 snapshots (see Appendix~\ref{sec:analytic_potentials}).
Any effects of dynamical friction of relatively low mass GCs are secondary to the variations in the merger history. 
We emphasize that the GC catalogs already matched the total mass, mass function, spatial distribution, velocity dispersion, and metallicity distribution with the MW GC population and surrounding dwarfs. These are the most important properties to reconstruct the statistics and observational characteristics of the GC stream populations.

For the tidal dissolution of GCs, we rely on the tidal tensor prescription described in Section~\ref{sec:sims}. 
However, \cite{Gieles2021} showed that populations of stellar mass black holes in GCs might play an important role in the formation and present day observability of GC streams. In fact, populations of stellar mass black holes in GCs can lead to much denser streams, which have peak densities closer to their dissolved clusters as compared to dissolved GCs without considering black holes \citep{roberts2024}. 

To fully forecast the missing population of observable GC streams in the MW and other galaxies, the impact of tidal shocks, populations of black holes in GCs, and the Milky Way foreground need to be implemented in more detail. In this paper, we have focused on forecasting how many streams there are in principle.

\subsection{Science with complete MW stream sample} \label{sec:futurescience}

A complete sample of stellar streams in the Milky Way with deeper photometry could vastly improve current stellar stream science. If we are indeed missing $\sim$1000 streams in the inner Galaxy, and $\sim$100 streams beyond $R_{\rm Gal}>15$~kpc, as our models predict, the detection of such streams can open up new interesting science directions.

Several studies have shown that modeling multiple streams simultaneously yields tighter constrains on the Galactic potential of the Milky Way. \citet{bovy2016} used the Palomar 5 \citep{odenkirchen2001} and GD-1 \citep{grillmair2006} stellar streams in the Milky Way to constrain the shape of the dark matter halo within 20 kpc of the Galactic center. \citet{bonaca2018} later demonstrated that GC streams on different orbits and locations in the Galaxy provide information on the local acceleration field of the halo. In particular, they found that longer, more eccentric streams contained more information about the dark matter halo.  
Recently, \citet{nibauer2022} showed that they could constrain the local acceleration field of  galaxies without any assumptions of the underlying potential if they treated the stream as a collection of orbits with a locally similar mixture of energies, rather than assuming that the stream delineates a single stellar orbit. In \citet{nibauer23}, they showed that the geometry of streams alone can be used to constrain potential shapes. Thus finding more streams or the fainter parts of streams that have already been detected can be used to further constrain the dark matter halo of the Milky Way at various galactocentric radii. 

Certain regions of the Milky Way's potential with overlapping orbit families, and resonances, can lead to deformations of streams in the form of stream-fanning \citep{pearson2015,pricewhelan2016a,sesar2016,pricewhelan2016b, yavetz2021}, bifurcations \citep{yavetz2023}, and splitting of streams \citep{Dodd2022}. As a consequence, the existence of thin, long streams in the Galactic halo tells us where regular orbits exist, and map out stable regions of the potential, which support the formation and coherence of thin streams over time. Having a complete picture of the regions in the halo that support such streams can therefore help us constrain the Galactic potential. 

The accretion of massive satellites such as the Sagittarius dwarf and the Large Magellanic Cloud can perturb stellar streams in the Milky Way \citep[see e.g.,][]{erkal2019, shipp2021,dillamore2022,Lilleengen2022,koposov2023} which affects constraints on the MW dark matter halo \citep{Vasiliev2021a,brooks24}. If we detect more streams in the outskirts of the MW in future data, perturbations to these streams can be used to constrain the orbits and mass distributions of in-fallen satellites. 
Thus, there are several exciting prospects for using streams for new science in addition to the indirect detections of subhalo interactions from gaps in streams as discussed in Section~\ref{sec:intro} \citep[see also][]{wagner2019}.

\subsection{External galaxies} \label{sec:discexternal}

The fact that the MW could have a multitude of undetected GC streams at large galactocentric radii is encouraging both for future MW surveys and for using streams for dark matter science in external galaxies \citep[see e.g.,][]{fardal2013,pearson2022b,nibauer23,aganze2023}. \citet{Pearson2019} showed that {\it Roman} will find GC streams with up to 5 times the initial mass of Pal 5 out to 6.2 Mpc for optimistic star--galaxy separation estimates. This volume includes $\sim$ 500 galaxies \citep{Karachentsev2019}.  When {\it Roman} observes external galaxies, stellar streams at large galactocentric radii will have a larger contrast in surface density to the galaxy stellar halos and such streams are therefore easier to detect. The work presented in this paper shows that it is likely that streams exist in the outskirts of MW-type galaxies.

Our closest massive neighbor, M31, has $\sim$3 times as many GCs as the MW with a more extended distribution. This is likely due to the more recent accretion events occurring in M31 as compared to the MW \citep{mackey2019}. 
Using a higher initial cluster mass cutoff for GCs of $10^5$ M$_{\odot}$, \cite{chengnedin2022} produced catalogs matching M31's GC distribution with a M31-like accretion history, mass profile, which matched the spatial distribution, kinematics, and  metallicities of M31's GCs. Specifically, we require them to have 1) total mass within the range $10^{12-12.5}\ \msun$ and 2) at least one major merger in the last 6~Gyr, resembling the likely progenitor of M32 \citep{souza2018}. Among the TNG50 galaxies that match these criteria, we focus on the one that best matches the number and metallicity distribution of M31 GCs. 

\begin{figure*}
	\includegraphics[width=\textwidth]{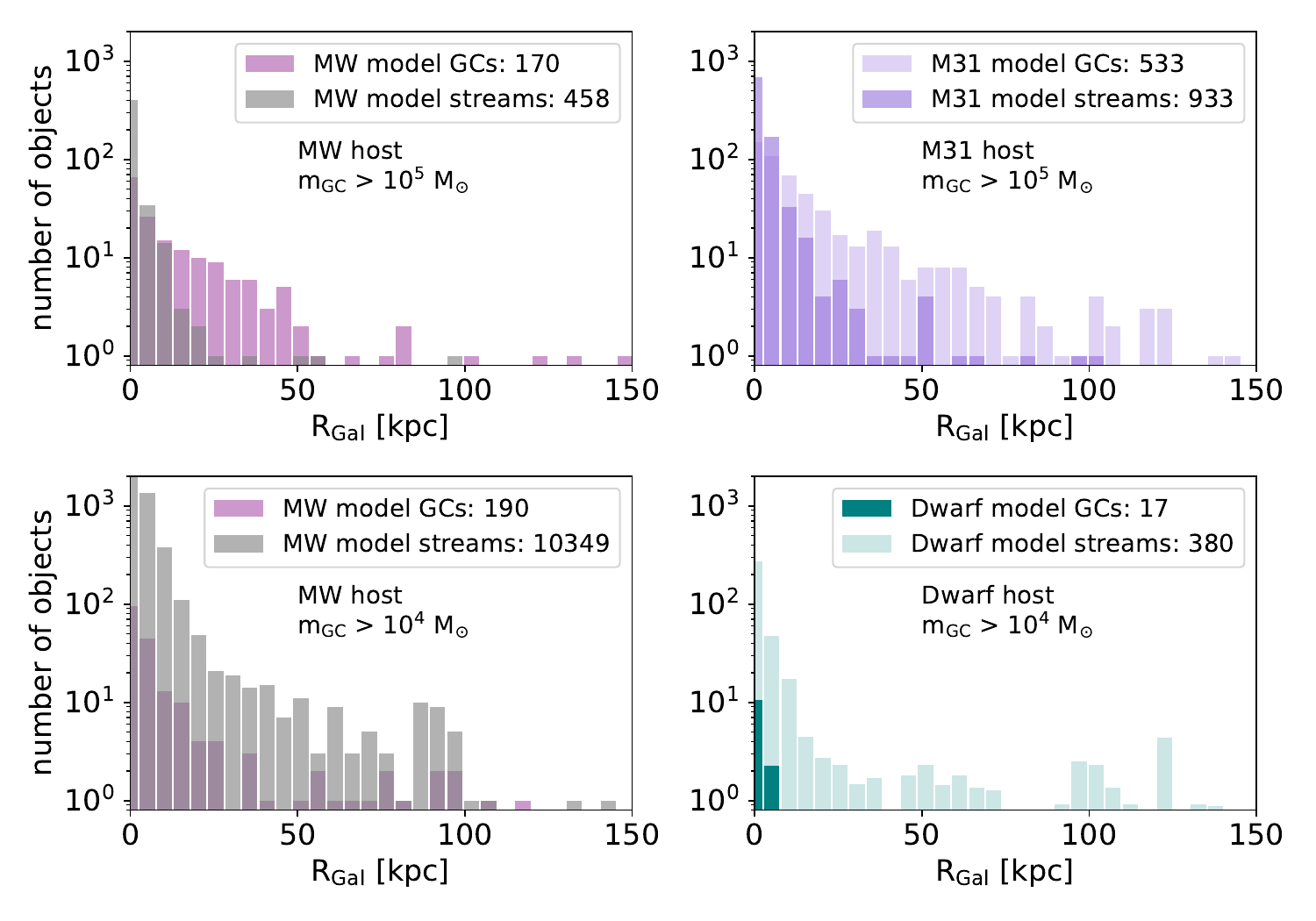}
    \caption{{\bf Top}: Comparison between the MW model's galactocentric distribution of GCs (purple) and fully disrupted GCs (gray) (left) and the best match M31 model's galactocentric distributions of GCs (light blue) to fully disrupted GCs (dark blue) (right) for catalogs including globular clusters with initial masses of m$_{\rm GC} > 10^5$ M$_{\odot}$. There is a large population of potential stream candidates at larger galactocentric distances in M31. 
    {\bf Bottom:} Comparison between MW1's galactocentric distributions of GCs (purple) and fully disrupted objects (gray) (left) and a model dwarf model's average galactocentric distributions of GCs (dark teal) to streams (light teal) for catalogs including globular clusters with initial masses of m$_{\rm GC} > 10^4$~M$_{\odot}$. There is a large number of predicted potential stream candidates in the dwarf model. 
    Note that the lower left distributions are the same as presented in Figure~\ref{fig:numberofobjects}.}
    \label{fig:M31}
\end{figure*}

In order to do a corresponding comparison to the M31 catalogs,
in Figure~\ref{fig:M31} (top, left) we show the number of MW model streams (gray) and GCs (pink) for the MW models with an initial cluster mass cutoff for GCs of $10^5$ M$_{\odot}$. In Figure~\ref{fig:M31} (top, right), we show the same plot for the best match M31 model's GCs (light purple) and stream distribution (dark purple) vs galactocentric radius. 

There is a larger population of GCs at larger galactocentric radii in M31 as compared to the MW, which is expected since the model catalogs reproduce the observed spatial distribution of M31 GCs. Additionally, there is an excess of fully disrupted GCs (stellar stream candidates) in the M31 catalog at  $\approx 30-60$ kpc from the center of the galaxy, as compared to the MW catalog with  the $10^5$ M$_{\odot}$ mass cutoff. Most of the  M31 streams candidates are located near the center of the galaxy, and there are only few streams beyond 60 kpc, likely since the tidal field is weak in the outskirts, which makes it hard for $10^5$~M$_{\odot}$ objects to tidally disrupt.

Note that we do not go through the steps of determining the survivability or observability of these M31 stream candidates, but based on the comparison to Section~\ref{sec:resobsstreams}, we conclude that there should be a substantial number of GC streams which can be observed with {\it Roman} in M31 \citep[see also][]{Pearson2019,Pearson2022}, revealing a new GC stream sample to explore. \citet{aganze2023} showed that {\it Roman} will be able to detect gaps in GC streams from interactions with dark matter subhalos out to $2-3$ Mpc, including M31. This is an exciting avenue to distinguish between dark matter candidates, as various candidates predict different subhalo mass functions and therefore stream gap distributions \citep{Bullock2017}. 

Most galaxies in the nearby Universe are dwarf galaxies \citep{Karachentsev2019}. Since {\it Roman} will only be able to detect GC streams in the nearby Universe \citep{Pearson2019,Pearson2022}, detecting GC streams in dwarfs would vastly expand the amount of host galaxies in which we could study GC streams.

There have been several observational confirmations of GCs in dwarf galaxies. 
The Exploration of Low-Mass VolumE Satellites Survey \citep[ELVES,][]{Carlsten2022a,Carlsten2022b} has explored the presence of GC systems in a sample of 140 confirmed early type dwarf satellite galaxies with stellar masses between $10^{5.5}-10^{8.5}$~M$_{\odot}$. They find that the occupation fraction of GCs significantly rises from 0 to 100\% for galaxies with stellar masses of $10^{6}-10^{8}$~M$_{\odot}$.
Several groups have also observed GCs along the streams of disrupted dwarf galaxies --- e.g., in the Giant Southern Stream in M31 (\citealt{Veljanoski2014}), along tidal debris in Centaurus A (\citealt{Hughes2022}), and along the Sagittarius stream in the Milky Way (\citealt{malhan2022}) --- which confirms that dwarfs indeed host their own population of GCs.  

The theoretical study of dwarf host galaxies in \citetalias{Chen2023} found that dwarfs with stellar masses $M_* >10^8$ M$_{\odot}$ contain GCs, with a median number $\sim$10 GCs per dwarf. To generate these catalogs of dwarf GCs, \citetalias{Chen2023} ran a suite of collisionless zoom-in simulations of the Local Group and investigated GC systems in dwarf satellites of the Milky Way and M31 analogs. They applied a GC formation model similar to the one employed in this work. However, they only selected dark matter particles near the galactic center as GC tracers, whereas we preferentially use stellar particles (see details in Section \ref{sec:sims}). GCs represented by these dark matter particles could still match the observed spatial and kinematic distributions as they were carefully selected in local density peaks to mimic the location of giant gas clouds. For these catalogs, \citetalias{Chen2023} used a GC mass cutoff of $10^4$ M$_{\odot}$. They run 25 realizations of each of the dwarf host galaxies of various masses. The catalogs also list the fully disrupted GCs in these dwarfs. 

In Figure~\ref{fig:M31} (bottom panels) we investigate the prospects for finding fully disrupted GCs in dwarf galaxies. We focus on 25 realizations of one dwarf host from \citetalias{Chen2023}, which has $M_* = 1.67 \times 10^8$~M$_{\odot}$ (halo ID = 2475241).
In order to compare the GC and stream candidate populations in the dwarf host to our findings from the MW, in Figure~\ref{fig:M31} (bottom left) we again show the number of MW model streams (gray) and model GCs (purple) from Figure~\ref{fig:numberofobjects}, which has a lower mass initial cluster mass cutoff for GCs of $10^4$~M$_{\odot}$ for MW1.
In Figure~\ref{fig:M31} (bottom right) we visualize the galactocentric radius distributions of the GCs (dark teal) and fully disrupted GCs in the dwarf host (light teal) averaged over the 25 realizations from \citetalias{Chen2023}. 

We find that there are several fully disrupted GCs in the outskirts of this dwarf host halo, which is exciting for future prospects of large samples of GC stream hosts observed with {\it Roman}.
In this work, we do not explore the observability and survivability of the potential streams candidates in dwarf hosts. 
GC disruption and the subsequent survival time of stellar streams in dwarfs remains largely unexplored \citep[although see][]{penarrubia2009, malhan:2021}. We conclude that GC stream populations could exist in dwarfs, if the streams remain observable after disruption.

\section{Conclusion} \label{sec:concl}

In this paper we have analyzed hierarchical models of globular cluster formation (based on the models presented in \citetalias{chengnedin2022} and \citetalias{Chen2023}) and simulated the evolution of the dissolved clusters to forecast the masses, numbers and radial distribution of stellar streams from globular clusters. Below, we summarize the key findings.

\begin{itemize}
    \item There are more than 10,000 fully disrupted GC objects with masses $>10^4$ M$_{\odot}$ in Milky Way-type galaxies. About $\sim 1000$ survive as coherent streams until present day. 

    \item Our results indicate that we have observed $<$10\% of the fully disrupted GC stream population in the Milky Way. Hundreds of missing streams reside within  $R_{\rm Gal} <30$~kpc, and tens of missing streams reside beyond $R_{\rm Gal} >30$~kpc.

    \item In the outskirts of MW type galaxies ($R_{\rm Gal} >30$~kpc), the population of surviving GC streams at present day are all from accreted GCs. 

    \item LSST will be able to detect even the most distant missing GC streams in resolved stars.  

    \item Our results indicate that GC streams should be abundant in the outskirts of M31 and nearby dwarf galaxies. These can be detected with future observations from the Nancy Grace Roman Space Telescope.
\end{itemize}

\acknowledgements
We thank the CCA Dynamics Group at the Flatiron Institute for insightful discussions. 
This work was initiated at the Aspen Center for Physics, which is supported by National Science Foundation grant PHY-2210452.
This project has received funding from the European Union's Horizon 2022 research and innovation programme under the Marie Skłodowska-Curie grant agreement No 101103129. This work was supported by a research grant (VIL53081) from VILLUM FONDEN. 
This research benefited from the Dwarf Galaxies, Star Clusters, and Streams Workshop hosted by the Kavli Institute for Cosmological Physics.

\software{
   \texttt{Astropy} ~\citep{astropy13,astropy18}, 
    ~\texttt{Matplotlib} ~\citep{Hunter:2007}, 
    ~\texttt{Gala} ~\citep{gala2017,gala2020}, 
    ~\texttt{Numpy} ~\citep{walt2011}, 
    ~\texttt{Scipy} ~\citep{scipy},
    ~\texttt{Galstreams} ~\citep{Mateu2023},
    ~\texttt{Agama} ~\citep{vasiliev2019}.
}

\appendix

\section{Analytic potential forms}
\label{sec:analytic_potentials}

In this work, the gravitational potentials of the TNG50 galaxies is represented by a combination of three analytic potentials for the bulge, the disk, and the dark-matter halo. We use the spherically symmetric \citet{hernquist:1990} profile for the bulge:
\begin{equation}
\Phi_{\rm bulge}(r) = -\frac{GM_b}{r + a_b}
\label{eq:hernquist}
\end{equation}
where $r$ is the galactocentric distance, $G$ is the gravitational constant, $M_b$ is the bulge mass, and $a_b$ is its scale radius.

We represent the stellar disks with an axisymmetric \citet{mn:1975} potential:
\begin{equation}
\Phi_{\rm disk}(R,z) = -\frac{GM_d}{\sqrt{R^2 + \left(a_d + \sqrt{z^2 + b_d^2}\right)^2}}
\label{eq:miyamoto-nagai}
\end{equation}
where $R$ is the cylindrical radius, $z$ is the distance from the disk mid-plane, $M_d$ is the disk mass, $a_d$ is the disk scale length, and $b_d$ is the disk scale height.

For the dark matter halo, we use a generalized NFW profile \citep{nfw:1996}:
\begin{align}
\Phi_{\rm halo}(R,z) &= - \frac{G M_h}{r_h} \frac{\ln\left(1 + u\right)}{u} 
  \label{eq:nfw} \\
u &= \frac{\sqrt{R^2 + (z/q)^2}}{r_h}
\end{align}
where $x,y,z$ are the Cartesian coordinates (with the $z=0$ being the disk plane), $M_h$ is the halo scale mass, $r_h$ is the halo scale radius, and $q$ is the flattening parameter.

To find the best-fitting parameters for the three components in the TNG50 galaxies at $z=0$, we first approximate the potential of the simulated galaxies using the \texttt{Agama} \citep{vasiliev2019} package. This package spline interpolates the numerical potential on a coordinate grid. The baryonic contribution to the potential, from all the gas and stellar particles bound to the galaxy, is assumed to be axisymmetric. We describe it as $\Phi_{\rm baryon, TNG50}(R,z)$ in a cylindrical coordinate system and evaluate it on a $20\times20$ grid evenly spaced in the $\log R$ and $\log z$ directions. The final baryonic potential is approximated using a 2D quintic spline on this grid. The dark matter potential is also axisymmetric but approximated by a spherical harmonic expansion in a spherical coordinate system, 
\begin{equation}
    \Phi_{\rm halo,TNG50}(r,\theta) = \sum_{l=0}^{l_{\rm max}}\Phi_l(r)\,Y_l^0(\theta)
    \label{eq:pot_halo}
\end{equation}
where $Y_l^0(\theta)$ are the axisymmetric terms of the spherical harmonics. The radius-dependent coefficients $\Phi_l(r)$ are calculated on 20 grid points evenly spaced in $\log r$ and connected via quintic splines. We find $l_{\rm max}=2$ suffices to approximate the value of the potential provided by the simulation output to $<2\%$ accuracy.

Given the \texttt{Agama} potentials, we obtain the fit parameters by minimizing the squared error loss function,
\begin{equation}
    \lambda(\Theta) = \sum_{r,\theta,\phi}\left[\Phi(r,\theta,\phi;\Theta)-\Phi_{\rm TNG50}(r,\theta,\phi)\right]^2,
    \label{eq:lost}
\end{equation}
where $\Theta=(M_b,a_b,M_d,a_d,b_d)$ or $\Theta=(M_h,r_h,q)$ are the parameters for Eq.~(\ref{eq:hernquist}) $+$ Eq.~(\ref{eq:miyamoto-nagai}) or Eq.~(\ref{eq:nfw}), respectively. The summation is over the grid of $(r,\theta,\phi)$ values: 20 grid points evenly spaced in the $\log r$ space for $r\in[0.1,1000]$~kpc, 20 grid points evenly spaced in $\theta\in[0,\pi]$, and 20 grid points evenly spaced in $\phi\in[0,2\pi)$. 

\begin{figure}
    \centering
    \includegraphics[width=\columnwidth]{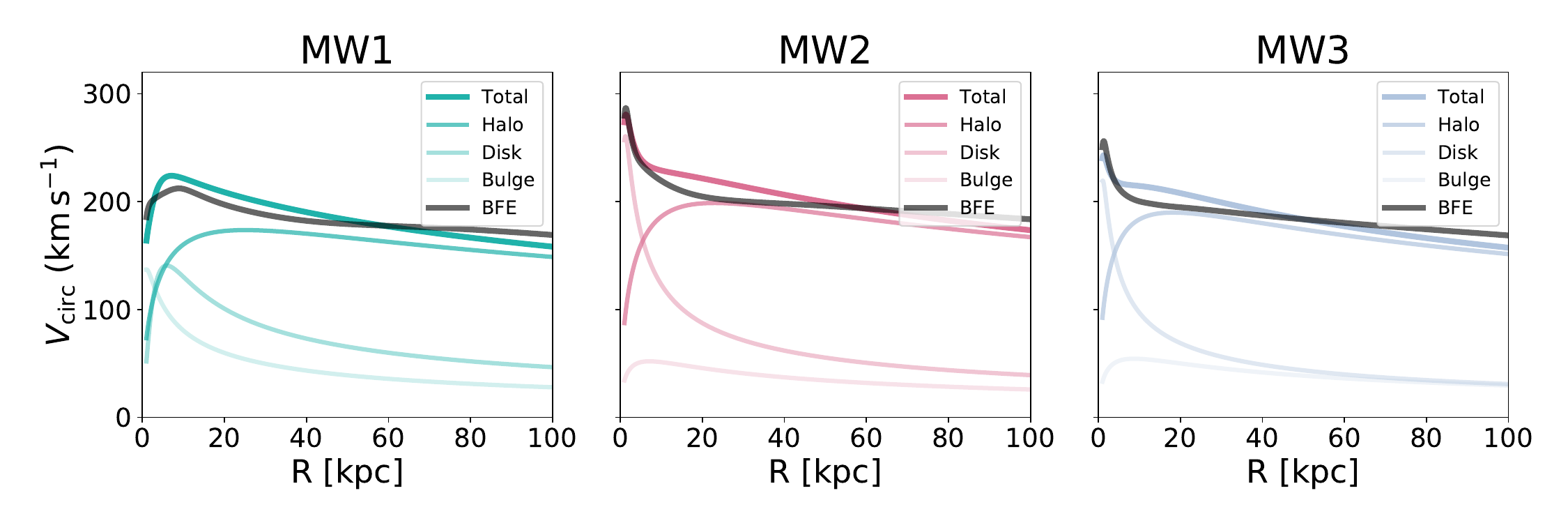}
    \caption{Comparison between the circular velocity for the BFE potential from TNG50 at present day (black) compared to the three-component analytical circular velocity curves for MW1 (left), MW2 (middle), and MW3 (right).}
    \label{fig:BFE}
\end{figure}

Figure~\ref{fig:BFE} shows the resulting decomposition of the potential into its three components. For example, for the TNG50 galaxy with halo ID \texttt{523889} (MW1) the fit parameters are
\begin{align}
    (M_b,a_b)&=(10^{10.26}\ \msun, 1.05\ \kpc) \\
    (M_d,a_d,b_d)&=(10^{10.70}\ \msun,  4.20\ \kpc, 0.006\ \kpc) \\
    (M_h,r_h,q)&=(10^{11.57}\ \msun, 11.6\ \kpc, 0.94)
\end{align}
The fit potential deviates from the actual simulation potential only by $\lesssim7\%$ within $r<200$~kpc. Such an error is small enough that we can ignore its influence on the calculation of the orbits.

\section{Failed mock streams}
\label{sec:failedstreams}

\begin{figure}
    \centering
    \includegraphics[width=\columnwidth]{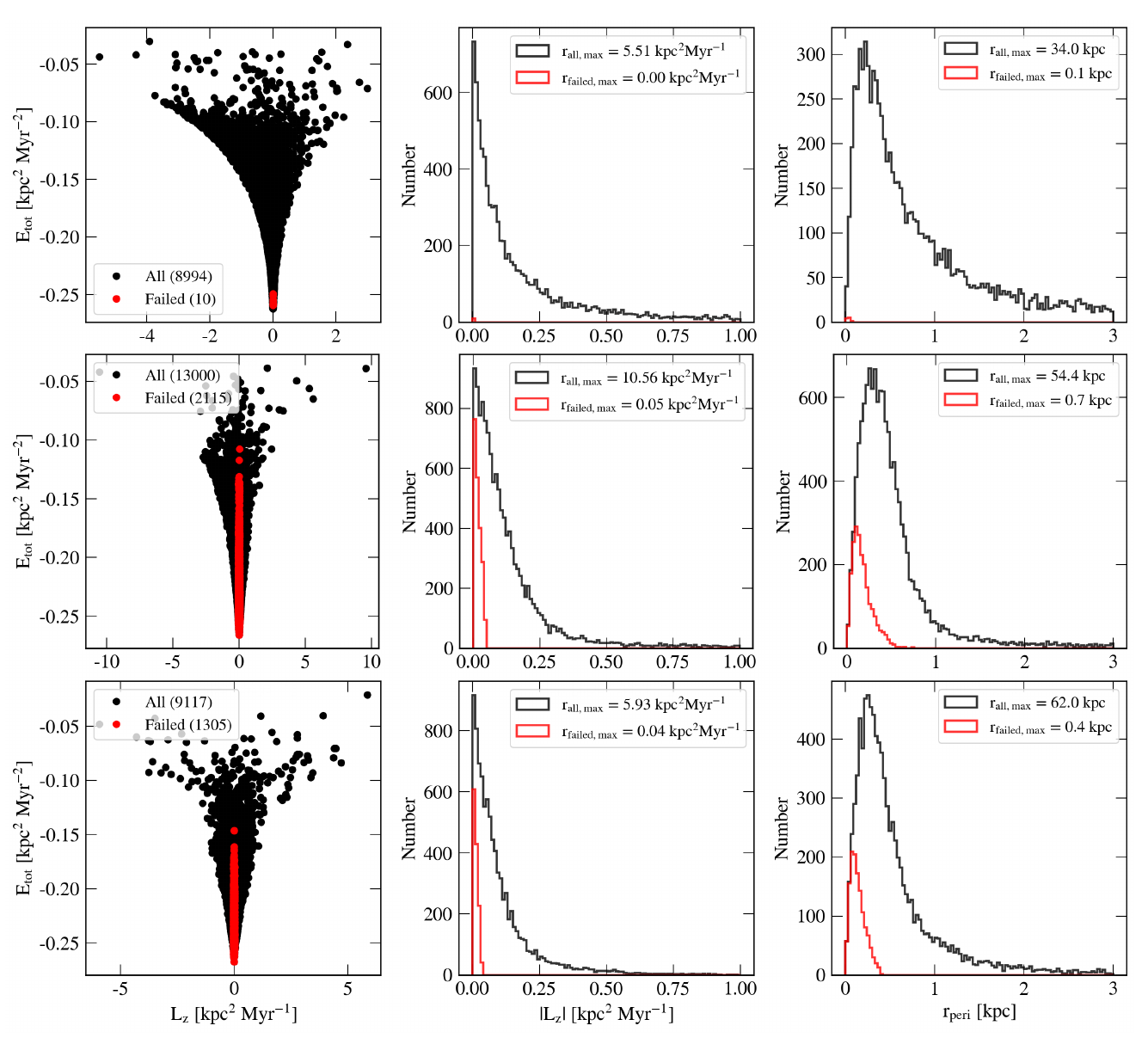}
    \caption{Orbital diagnostics of all fully disrupted objects from the MW1 catalog (\textit{top row}), MW2 catalog (\textit{middle row}), and MW3 catalog (\textit{bottom row}). \textit{Left panels:} the total energy of each fully disrupted object (black) and each failed mock stream (red) as a function of its $z$-angular momentum. \textit{Middle panels:} The number of objects as a function of their absolute value of their  $z$-angular momentum. \textit{Right panels:} The number of objects as a function of their orbital pericenter. MW1 has much fewer failed objects, and fewer fully disrupted objects in the central region of the potential as compared to MW2 and MW3.}
    \label{fig:failed}
\end{figure}

For each of the three MW analogs, we simulate streams for all mock GCs that remained bound at accretion and then fully disrupted at present. However, a subset for mock streams failed. To compare the properties of the failed objects to the properties of the mock streams that did not fail, in Figure~\ref{fig:failed} we summarize the orbital characteristics for all objects (black) as well as for the failed objects (red). For each MW analog, the mock stream objects which fail have a $z$-component of the angular momentum close to zero, and thus a small net rotation around the disk (left middle panel). These objects also all have small pericenters ($r_{\rm peri}<0.7$ kpc, right panel). We note that two of the TNG50 galaxies, MW2 and MW3, have very extended spheroids (``bulges'') but small disks. This creates numerical issues for the stream generator, but only for the in-situ clusters.

\end{document}